\documentclass[aps,prb,reprint,twocolumn,showpacs,preprintnumbers,superscriptaddress]{revtex4-2}
\usepackage{graphicx}
\usepackage{amsmath}
\usepackage{amsthm}
\usepackage{bm}
\usepackage{float}
\usepackage{txfonts}
\usepackage{amsfonts}
\usepackage{amssymb}
\usepackage{mathtools}
\usepackage{array}
\usepackage{ulem}
\usepackage{multirow,bigdelim}
\graphicspath{{./figures/}}
\newcommand{\Reimei}{{\rm Reimei}}

\usepackage[colorlinks=true,citecolor=blue,urlcolor=blue]{hyperref}

\begin{document}

\preprint{RIKEN-iTHEMS-Report-26}

\title{Dissipative ground-state preparation of a quantum spin chain on a trapped-ion quantum computer}
\author{Kazuhiro~Seki}
\email{kazuhiro.seki@riken.jp}
\affiliation{Quantum Computational Science Research Team, RIKEN Center for Quantum Computing (RQC), Saitama 351-0198, Japan}

\author{Yuta Kikuchi}
\email{yuta.kikuchi@quantinuum.com}
\affiliation{Quantinuum K.K., Otemachi Financial City Grand Cube 3F, 1-9-2 Otemachi, Chiyoda-ku, Tokyo, Japan}
\affiliation{RIKEN Center for Interdisciplinary Theoretical and Mathematical Sciences (iTHEMS), RIKEN, Wako 351-0198, Japan}

\author{Tomoya Hayata}
\email{hayata@keio.jp}
\affiliation{Department of Physics, Keio University School of Medicine, 4-1-1 Hiyoshi, Kanagawa 223-8521, Japan}
\affiliation{RIKEN Center for Interdisciplinary Theoretical and Mathematical Sciences (iTHEMS), RIKEN, Wako 351-0198, Japan}
\affiliation{International Center for Elementary Particle Physics and The University of Tokyo, 7-3-1 Hongo, Bunkyo-ku, Tokyo 113-0033, Japan}

\author{Seiji Yunoki}
\email{yunoki@riken.jp}
\affiliation{Quantum Computational Science Research Team, RIKEN Center for Quantum Computing (RQC), Saitama 351-0198, Japan}
\affiliation{Computational Quantum Matter Research Team, RIKEN Center for Emergent Matter Science (CEMS), Wako, Saitama 351-0198, Japan}
\affiliation{Computational Materials Science Research Team,
RIKEN Center for Computational Science (R-CCS), Kobe, Hyogo 650-0047, Japan}
\affiliation{Computational Condensed Matter Physics Laboratory,
RIKEN Pioneering Research Institute (PRI), Saitama 351-0198, Japan}

\date{\today}

\begin{abstract}
We demonstrate a dissipative protocol for ground-state preparation of a quantum spin chain on a trapped-ion quantum computer. 
As a first step, we derive a Kraus representation of a dissipation channel for the protocol recently proposed by Ding {\it et al}. [\href{https://link.aps.org/doi/10.1103/PhysRevResearch.6.033147}{Phys. Rev. Res. {\bf 6}, 033147 (2024)}] that still holds for arbitrary temporal discretization steps, extending the analysis beyond the Lindblad dynamics regime. 
The protocol guarantees that the fidelity with the ground state monotonically increases (or remains unchanged) under repeated applications of the channel to an arbitrary initial state, provided that the ground state is the unique steady state of the dissipation channel. 
Using this framework, we implement dissipative ground-state preparation of a transverse-field Ising chain for up to 19 spins on the trapped-ion quantum computer \Reimei{} provided by Quantinuum. 
Despite the presence of hardware noise, the dynamics consistently converges to a low-energy state far away from the maximally mixed state even when the corresponding quantum circuits contain as many as 4110 entangling gates, demonstrating the intrinsic robustness of the protocol. 
By applying zero-noise extrapolation, the resulting energy expectation values are systematically improved to agree with noiseless simulations within statistical uncertainties. 
\end{abstract}\maketitle

\section{Introduction}

Estimating ground-state properties, such as energies, correlations, and entanglement, remains one of the most important and challenging tasks in quantum many-body physics~\cite{Wu2024}. 
A broad range of classical numerical methods, including exact diagonalization, variational Monte Carlo, quantum Monte Carlo, and density-matrix-renormalization group techniques, have been developed and benchmarked against paradigmatic models of strongly correlated systems, such as the Hubbard models~\cite{Hirsch1985,Tahara2008,LeBlanc2015,Zheng2017,Nomura2017,Ido2018,Seki2019,Sorella2023} and the Heisenberg models~\cite{Sandvik1997,Iqbal2013,Iqbal2016,Nomura2021}. These approaches have yielded valuable insights and cross-validations; however, accurately determining the ground-state properties of these systems remains 
classically intractable for large system sizes, especially in spatial dimensions greater than one. 
Recent quantum resource analyses suggest that ground-state energy estimation of the Hubbard and Heisenberg models via quantum phase estimation could offer one of the earliest demonstrations of practical quantum advantage once fault-tolerant quantum computers become available~\cite{Yoshioka2024}.
Ground-state energy estimation of molecules is likewise considered a promising target for demonstrating quantum advantage~\cite{lanes2025frameworkquantumadvantage}.

Despite such a promising assessment, quantum phase estimation will likely require fault-tolerant quantum computers to address classically intractable problems. 
In the meantime, the variational quantum eigensolver (VQE) has been widely used to study ground-state properties of quantum many-body systems on noisy intermediate-scale quantum (NISQ) devices~\cite{Peruzzo2014,Kandala2017}. 
Although VQE suffers from the notorious barren-plateau problem when optimizing variational parameters in deep circuits~\cite{McClean2018}, it has nevertheless served as a cornerstone of quantum-classical hybrid algorithms. 
Recently, the quantum-selected configuration interaction (QSCI) method~\cite{Kanno2023}, also referred to as the sample-based quantum diagonalization (SQD) method~\cite{Robledo_Moreno2025}, has been proposed as an alternative ground-state preparation protocol suitable for NISQ devices. 
The QSCI/SQD method avoids the variational-parameter optimization of VQE and simultaneously mitigates the measurement overhead associated with estimating energy expectation values. 
Following its initial proposal, several variants of the method have been developed~\cite{Nakagawa2024,Sugisaki2025,Mikkelsen2025,Yu2025} and implemented in the context of quantum-centric supercomputing~\cite{Alexeev2024} for large-scale electronic structure calculations using state-of-the-art quantum computers and classical supercomputers~\cite{Robledo_Moreno2025,Barison2025, Shirakawa2025}.
As with many emerging quantum algorithms, however, the superiority of QSCI/SQD over classical selected-CI approaches, such as the heat-bath CI method~\cite{Holmes2016}, remains under active discussion~\cite{Barison2025,Reinholdt2025}, and further development of QSCI/SQD is ongoing.

The capability of quantum devices is growing rapidly. In particular, the ability to perform mid-circuit measurement and reset (MCMR) has not only enabled the simulation of open-quantum-system dynamics~\cite{Koh2023,Chertkov2023}, but also opened new avenues for exploring dissipation in optimization and state preparation~\cite{FossFeig2021,DeCross2022}. 
For example, MCMR makes it possible to prepare thermal Gibbs states on a quantum computer using Lindblad or open-quantum-system dynamics~\cite{Shtanko2023,chen2023quantum,chen2023efficient,Brunner2025,lloyd2025quantumthermalstatepreparation,ding2025endtoendefficientquantumthermal}. 
Beyond thermal-state preparation, dissipative protocols for ground-state preparation have recently attracted significant attention~\cite{Mi2024,Ding2024,Langbehn2024,Matthies_2024,Lambert2024,langbehn2026universalcoolingquantumsystems} (see, e.g., Ref.~\cite{Lin:2025rzs} for a review).
One example is the protocol demonstrated on a superconducting quantum processor by Mi {\it et al.}~\cite{Mi2024}, in which quasiparticle excitations are released from the system to the environment, enabling efficient preparation of the ground state of one-dimensional (1D) transverse-field Ising models.
More generally, Ding {\it et al.} proposed a protocol for preparing ground states of arbitrary quantum many-body Hamiltonians based on a completely positive and trace-preserving (CPTP) map specifically engineered to have the ground state as its unique steady state~\cite{Ding2024}.
While this protocol is envisioned as an early-fault-tolerant quantum algorithm, it is important to examine its implementation and performance on currently available noisy quantum devices.

In this paper, we demonstrate a dissipative ground-state preparation protocol for a 1D transverse-field Ising model using the trapped-ion quantum computer \Reimei{}~\cite{reimei_benchmark}.
First, we derive a Kraus representation of the dissipation channel that remains valid for arbitrarily large temporal-discretization steps. 
We then show that the fidelity with respect to the ground state is monotonically non-decreasing under repeated applications of the channel to an arbitrary initial state. 
We implement the protocol on \Reimei{} to estimate the ground-state energy of the transverse-field Ising model for system sizes of up to 19 spins using a single ancilla qubit. 
By adopting zero-noise extrapolation (ZNE), we systematically obtain noise-mitigated energy expectation values that are consistent with noiseless simulations within statistical uncertainties.

The rest of the paper is organized as follows. 
In Sec.~\ref{sec:method}, we review the dissipative ground-state preparation protocol proposed in Ref.~\cite{Ding2024} and 
reformulate it by deriving a Kraus representation of the corresponding CPTP map. 
In Sec.~\ref{sec:results}, we present experimental results for dissipative ground-state preparation of a spin chain on a trapped-ion quantum device. 
In Sec.~\ref{sec:discussion}, we summarize our findings and discuss future perspectives. 
Additional details of the formalism, classical numerical results, and experimental results are provided in the Appendices.

\section{Method}\label{sec:method}

In this section, we review the dissipative ground-state preparation protocol proposed in Ref.~\cite{Ding2024}.
We reformulate the protocol by deriving a Kraus representation of the corresponding dissipative channel and show, using the monotonicity of fidelity under CPTP maps, that the ground state is reached irrespective of the step size~$\tau$ of the associated dilated time evolution $\hat{\cal W}(\sqrt{\tau})$.

\subsection{Dissipative ground state preparation}

We aim to prepare the ground state of the Hamiltonian
\begin{equation}
\hat{H}=\sum_{i=0}^{d-1}E_i|E_i\rangle\langle E_i|, 
\end{equation}
where the eigenvalues satisfy  
%$E_0 < E_1 \leqslant E_2 \leqslant \cdots \leqslant E_{d-1}$, where $d$ is the dimension of the Hilbert space. 
$E_0=E_1=\cdots=E_{g_0-1}< E_{g_0}\leqslant E_{g_0+1} \leqslant \cdots \leqslant E_{d-1}$, with $g_0$ denoting the ground-state degeneracy. 
The Hilbert-space dimension is $d$, and  
$|E_0\rangle, |E_1\rangle,\cdots, |E_{g_0-1}\rangle$ span the degenerate ground-state manifold.
For notational convenience, we hereafter denote the ground state by $|E_0\rangle$; 
however, the following arguments apply equally to all degenerate ground states $|E_1\rangle, \cdots, |E_{g_0-1}\rangle$.
We assume that $\hat{H}$ acts on a system of $N$ qubits, implying $d=2^N$.

We introduce a dissipative dynamics described by a CPTP map $\Gamma_K$, such that the ground state $|E_0\rangle \langle E_0|$ is its unique steady state: 
\begin{equation}
\Gamma_K[|E_0\rangle\langle E_0|]=|E_0\rangle \langle E_0|.
\label{eq:K_fixed_point}
\end{equation}
Here, the subscript $K$ denotes a jump operator that will be introduced below.   
To construct $\Gamma_K$, we introduce an ancilla qubit and define a dilated unitary acting on the combined ancilla-system Hilbert space (i.e., $N+1$ qubits) as 
\begin{equation}
\hat{\cal W}(\sqrt{\tau}) 
= {\exp}(-{\rm i} \hat{\cal K}\sqrt{\tau}),
\label{eq:W}
\end{equation}
where $\tau$ is a positive real parameter, and $\hat{\cal K}$ is the dilated jump operator,
\begin{align}
    \hat{\cal K} 
& \equiv \begin{bmatrix}
 0 & \hat{K}^\dagger \\
 \hat{K} & 0
 \end{bmatrix}
  = |1\rangle \langle 0|_{\rm a} \otimes \hat{K} 
  +  |0\rangle \langle 1|_{\rm a} \otimes \hat{K}^\dagger. 
  \label{eq:calK} 
\end{align}
The jump operator $\hat{K}$, acting only on the system qubits, is designed to annihilate the ground state: 
\begin{equation}
\hat{K}|E_0\rangle=0.
\label{eq:Klambda0}
\end{equation}
An explicit construction of $\hat{K}$ will be given in Sec.~\ref{subsec:jump}.
% Notice that $\hat{K}$ is nonhermitian.
We also note that $\hat{\cal K}$ annihilates the product state $|0\rangle\langle0|_{\rm a}\otimes|E_0\rangle\langle E_0|$; that is, $\hat{\cal K}(|0\rangle\langle0|_{\rm a}\otimes|E_0\rangle\langle E_0|)\hat{\cal K}^\dagger=0$.

Suppose that the initial state of the total system is $|0\rangle\langle0|_{\rm a}\otimes \hat{\rho}$. 
The state evolved under $\hat{\cal W}(\sqrt{\tau})$ is then  
\begin{equation}
\hat{\sigma}(\tau)\equiv 
\hat{\cal W}(\sqrt{\tau})
\left(|0\rangle\langle0|_{\rm a}\otimes \hat{\rho}\right)
\hat{\cal W}^\dagger(\sqrt{\tau}).
\label{eq:sigmatau}
\end{equation}
We define the dissipative channel as 
${\Gamma}_{K}[\hat{\rho}]\equiv
{\rm Tr}_{\rm a}\left[\hat{\sigma}(\tau)\right]$.
By substituting the series expansion of $\hat{\cal W}(\sqrt{\tau})$ into Eq.~\eqref{eq:sigmatau} and tracing out the ancilla, we obtain 
(see Appendix~\ref{app:derivation_Kraus} for details) 
\begin{align}
{\Gamma}_{K}[\hat{\rho}]=
{\rm Tr}_{\rm a}\left[\hat{\sigma}(\tau)\right]
=
\hat{C} 
\hat{\rho} 
\hat{C}
+
\tau
\hat{K}
\hat{S}_{\rm c} 
\hat{\rho}
\hat{S}_{\rm c}
\hat{K}^\dagger, 
\label{eq:KrausCS}
\end{align}
where 
\begin{align}
\hat{C} & \equiv \cos\!{\sqrt{\tau \hat{K}^\dagger\hat{K}}}=\sum_{k=0}^{\infty} \frac{(-1)^k}{(2k)!}\left(\tau \hat{K}^\dagger\hat{K}\right)^{k},
\label{eq:Cos}\\
\hat{S}_{\rm c} & \equiv {\rm sinc}\!\sqrt{\tau \hat{K}^\dagger\hat{K}} = \sum_{k=0}^{\infty} \frac{(-1)^k}{(2k+1)!}\left(\tau \hat{K}^\dagger\hat{K}\right)^{k},
\label{eq:Sinc}
\end{align}
and $\sqrt{\hat{K}^\dagger\hat{K}}$ is well defined because $\hat{K}^\dagger \hat{K}$ is positive semidefinite. 
Introducing $\hat{M}_0=\hat{C}$ and $\hat{M}_1=-{\rm i}\sqrt{\tau}\hat{K}\hat{S}_c$, Eq.~\eqref{eq:KrausCS} can be expressed in the standard Kraus form:  
\begin{equation}
    {\Gamma}_{K}[\hat{\rho}]
    =
    \hat{M}_0
    \hat{\rho} 
    \hat{M}_0^\dagger
    +
    \hat{M}_1
    \hat{\rho} 
    \hat{M}_1^\dagger
\label{eq:KrausMs}
\end{equation}
with the completeness relation
\begin{equation}
\hat{M}_0^\dagger\hat{M}_0 + \hat{M}_{1}^\dagger\hat{M}_1 
= \cos^2\!{\sqrt{\tau \hat{K}^\dagger\hat{K}}}
+ \sin^2\!{\sqrt{\tau \hat{K}^\dagger\hat{K}}}
= \hat{I}.
\label{eq:Completeness}
\end{equation}
Thus, Eq.~(\ref{eq:KrausCS}) provides a Kraus (Stinespring) representation of the channel ${\Gamma}_K$.
Importantly, Eq.~(\ref{eq:K_fixed_point}) holds for arbitrary $\tau$ as long as the jump operator satisfies \(\hat{K}|E_0\rangle = 0\) [Eq.~\eqref{eq:Klambda0}]. 
As shown in Appendix~\ref{app:derivation_Kraus}, the dissipative dynamics generated by ${\Gamma}_{K}$ reduces to a Lindblad dynamics with jump operator $\hat{K}$ in the limit $\tau\to0$, up to $O(\tau^2)$.

We remark that the fidelity between the evolved state and the ground state is monotonically non-decreasing under the dissipative dynamics. 
For an arbitrary CPTP map ${\cal E}$ and density operators $\hat{\rho}$ and $\hat{\sigma}$, the fidelity satisfies 
%\begin{equation}
$
F({\cal E}[\hat{\rho}],{\cal E}[\hat{\sigma}]) \geqslant F(\hat{\rho},\hat{\sigma})$,
%\end{equation}
where the fidelity is defined as $F(\hat{\rho},\hat{\sigma})={\rm Tr}\sqrt{
\sqrt{\hat{\sigma}} 
\hat{\rho}
\sqrt{\hat{\sigma}}}$. 
Suppose that $\hat{\rho}^*$ is the unique steady state of ${\cal E}$, i.e., ${\cal E}[\hat{\rho}^*]=\hat{\rho}^*$.
Then, ${\cal E}[\hat{\rho}]=\hat{\rho}$ holds if and only if $\hat{\rho}=\hat{\rho}^*$. 
Furthermore, 
$
F({\cal E}[\hat{\rho}],{\cal E}[\hat{\rho}^*]) =
F({\cal E}[\hat{\rho}],\hat{\rho}^*) \geqslant F(\hat{\rho},\hat{\rho}^*)
$, 
where equality holds only when $\hat{\rho}=\hat{\rho}^*$. 
Replacing $\hat{\rho}$ with ${\cal E}^m[\hat{\rho}]$, we obtain
\begin{equation}
F({\cal E}^{m+1}[\hat{\rho}],\hat{\rho}^*) \geqslant F({\cal E}^{m}[\hat{\rho}],\hat{\rho}^*),
\label{eq:monotone_m}
\end{equation}
indicating that ${\cal E}^{m}[\hat{\rho}]$ monotonically approaches the unique steady state $\hat{\rho}^*$ as $m$ increases. 
In the present protocol, the CPTP map is the dissipative dynamics ${\cal E}=\Gamma_K$.
Reference~\cite{Ding2024} shows that the ground state $\hat{\rho}^*=|E_0\rangle\langle E_0|$ is the unique steady state of $\Gamma_{K}$ under appropriate assumptions on the jump operator $\hat{K}$ and the initial state.  
Given the uniqueness of the steady state, the ground-state fidelity is therefore monotonically non-decreasing under successive applications of $\Gamma_K$. 
We note, however, that monotonicity of the fidelity does not necessarily imply monotonicity of the energy. 
Table~\ref{table:1} highlights the characteristics of the dissipative dynamics method for ground-state preparation in comparison with the imaginary-time evolution.

\begin{table*}
\caption{
\label{table:1}
Comparison between the dissipative dynamics method and the imaginary-time evolution for ground-state preparation.
}
\begin{tabular}{c|ccc}
\hline
\hline
Method     & Operations & Monotonicity  & Time to convergence \\
\hline
Dissipative dynamics & Unitary $+$ Partial trace & Non-decreasing fidelity in time & $(\text{Liouvillian gap})^{-1}$ 
\\
Imaginary-time evolution    & Unitary $+$ Post selection & Non-increasing energy in time & $(\text{Energy gap})^{-1}$ 
\\
\hline
\hline
\end{tabular}
\end{table*}

\subsection{Jump operator and filter function}
\label{subsec:jump}

To obtain an explicit form of the jump operator, we write it as 
\begin{align}
\hat{K}
&=\sum_{i,j=0}^{d-1} \tilde{f}(E_i-E_j)|E_i\rangle\langle E_i|\hat{A}|E_j\rangle\langle E_j| 
\label{eq:Kw}, 
\end{align}
where $\hat{A}$ is an arbitrary operator and  $\tilde{f}$ is a filter function that satisfies 
\begin{equation}
\tilde{f}(\omega)=0 \quad \text{for} \quad \omega>0 
\label{eq:fwcond}
\end{equation}
so that $\hat{K}$ induces no energy-increasing transitions. 
Although we assume that $\hat{A}$ is Hermitian, the jump operator $\hat{K}$ is in general non-Hermitian.

For convenience in implementation, we rewrite the jump operator in the form of an operator Fourier transform (OFT)~\cite{chen2023quantum}:
\begin{align}
\hat{K}
= \int_{-\infty}^{\infty} {\rm d}s f(s) \hat{A}(s)
\quad\text{with}\quad
\hat{A}(s) = {\rm e}^{{\rm i}\hat{H}s}\hat{A} {\rm e}^{-{\rm i}\hat{H}s}.
\label{eq:Kt}
\end{align}
The filter function in the frequency domain, $\tilde{f}(\omega)$, is related to its time-domain representation $f(s)$ via 
%$\tilde{f}(\omega) = \int_{-\infty}^{\infty} {\rm d}s f(s){\rm e}^{{\rm i}\omega s}$, or equivalently, 
\begin{equation}
f(s)=\frac{1}{2\pi}\int_{-\infty}^{\infty}{\rm d}\omega\tilde{f}(\omega){\rm e}^{-{\rm i}\omega s}.
\label{eq:fs}
\end{equation}

\subsection{Interaction between ancilla and system}

To construct the Stinespring dilation unitary $\hat{\cal W}(\sqrt{\tau})$ as a quantum circuit, we rewrite the interaction operator $\hat{\cal K}$ between the system and the ancilla in Eq.~(\ref{eq:calK}) as 
\begin{align}
    \hat{\cal K} &
 = 
 \begin{bmatrix}
    0 & {\rm Re}\hat{K}-{\rm i}\,{\rm Im}\hat{K} \\
    {\rm Re}\hat{K}+{\rm i}\,{\rm Im}\hat{K}& 0
 \end{bmatrix}
 = \hat{X}_{\rm a} \otimes  {\rm Re}\hat{K} 
  +\hat{Y}_{\rm a} \otimes {\rm Im}\hat{K}, 
\end{align}
where $\hat{X}_{\rm a}$ and $\hat{Y}_{\rm a}$ are Pauli $X$ and $Y$ operators acting on the ancilla qubit, and 
${\rm Re}\hat{K}$ and ${\rm i}\, {\rm Im} \hat{K}$ denote the Hermitian and skew-Hermitian parts of $\hat{K}$:
\begin{align}
&{\rm Re}\hat{K} \equiv \frac{\hat{K}+\hat{K}^\dagger}{2} = \int_{-\infty}^{\infty} {\rm d}s\, {\rm Re} f(s) \hat{A}(s),\\ 
&{\rm Im}\hat{K} \equiv \frac{\hat{K}-\hat{K}^\dagger}{2{\rm i}} =  \int_{-\infty}^{\infty} {\rm d}s\, {\rm Im} f(s) \hat{A}(s). 
\end{align} 
%By rewriting the real and imaginary parts of $f(s)$ as 
Using ${\rm Re}f(s)=|f(s)|\cos\varphi(s)$ and ${\rm Im}f(s)=|f(s)|\sin\varphi(s)$ with $\varphi(s)=\arg{f(s)}$, we obtain
\begin{align}
\hat{\cal K}
= 
\int_{-\infty}^{\infty} {\rm d}s\,|f(s)|
\left(\cos{\varphi(s)}\hat{X}_{\rm a}+\sin{\varphi(s)}\hat{Y}_{\rm a}\right)\otimes\hat{A}(s).
\label{eq:calKs}
\end{align}
For convenience, we express the single-qubit operator acting on the ancilla in Eq.~(\ref{eq:calKs}) as   
\begin{equation}
\hat{P}(s) \equiv \cos{\varphi(s)}\hat{X}_{\rm a}+\sin{\varphi(s)}\hat{Y}_{\rm a} = \hat{R}_{Z_{\rm a}}(\varphi(s))\hat{X}_{\rm a}\hat{R}_{Z_{\rm a}}^\dagger(\varphi(s)), 
\label{eq:Ps}
\end{equation}
where $\hat{R}_{{Z}_{\rm a}}(\varphi)={\rm e}^{-{\rm i}\varphi\hat{Z}_{\rm a}/2}$ is a single-qubit $Z$-rotation acting on the ancilla.
The dilated unitary operator can then be expressed as 
\begin{equation}
\hat{\cal W}(\sqrt{\tau})=\exp\left[-{\rm i}\sqrt{\tau}\int_{-\infty}^{\infty} {\rm d}s\,|f(s)|
\hat{P}(s)\otimes \hat{A}(s)\right].
\label{eq:Wsa}
\end{equation}

\subsection{Truncation and discretization of the OFT}
\label{sec:TD_OFT}

We implement the unitary $\hat{\cal W}(\sqrt{\tau})$ by truncating and discretizing the time integral of the OFT in Eq.~\eqref{eq:Kt}, and by decomposing the resulting exponential using suitable Trotter steps. 
As we shall see later, we choose $f(s)$ such that $|f(s)|$ decays exponentially in $|s|$.  
We may therefore truncate the integral to the finite interval $[-S_s,S_s]$ and discretize it as 
\begin{align}
\hat{\cal K} 
\approx
\sum_{l=-M_s}^{M_s} \Delta_s |f(s_l)|
\hat{P}(s_l)
\otimes\hat{A}(s_l),
\label{eq:calK_discrete}
\end{align}
where $s_l = l\Delta_s$ ($l=-M_s, -M_s+1,\cdots,M_s$), and  $\Delta_s=S_s/M_s$ with $M_s$ being a positive integer. 
Under this time discretization, we use the second-order Trotter formula (with respect to $\Delta_s$) to approximate the unitary in Eq.~(\ref{eq:Wsa}) as 
\begin{equation}
\hat{\cal W}(\sqrt{\tau}) 
=
\prod_{l=-M_s}^{M_s}\hat{\cal U}(s_l)
\prod_{l=M_s}^{-M_s}\hat{\cal U}(s_l) + O(\Delta_s^3),  
\label{eq:Wdiscrete}
\end{equation}
where
\begin{align}
\hat{\cal U}(s_l)
&={\exp}\left[{-{\rm i}\frac{\sqrt{\tau}}{2} \Delta_s |f(s_l)| \hat{P}(s_l) \otimes \hat{A}(s_l)}\right] \notag 
\\
&\hspace{-.7em}\overset{\text{Eq.}\eqref{eq:Ps}}{=}
\hat{\cal V}^\dagger(s_l)
{\exp}\left[{-{\rm i}\frac{\sqrt{\tau}}{2} \Delta_s |f(s_l)| \hat{X}_{\rm a} \otimes \hat{A}}\right]
\hat{\cal V}(s_l)
\label{eq:Usl}
\end{align}
with
\begin{equation}
\hat{\cal V}(s_l)=
\hat{R}_{Z_{\rm a}}^\dagger(\varphi(s_l))\otimes {\rm e}^{-{\rm i}\hat{H}s_l}.
\end{equation}
% In Eq.~\eqref{eq:Usl}, we separate time dependencies of $\hat{A}(s_l)$ and $\hat{P}(s_l)$ from the entangling operation between the system and ancilla (the unitary at the middle) by using Eq.~\eqref{eq:Ps}.
%
Using $s_{l+1}=s_l+\Delta_s$, we further obtain 
\begin{equation}
\hat{\cal V}^\dagger(s_{l+1})
\hat{\cal V}(s_{l})
=
\hat{R}_{Z_{\rm a}}(\varphi(s_{l+1})) 
\hat{R}_{Z_{\rm a}}^\dagger(\varphi(s_{l}))
\otimes {\rm e}^{{\rm i}\hat{H}\Delta_s}.
\label{eq:VdagV}
\end{equation}
This shows that, in two consecutive applications $\hat{\cal U}(s_{l+1})\hat{\cal U}(s_l)$, the net operation on the system is the short-time evolution $\hat{I}\otimes{\rm e}^{{\rm i}\hat{H}\Delta_s}$ except at the first and last time slices. 
Using Eqs.~(\ref{eq:Usl})-(\ref{eq:VdagV}), the first term in Eq.~(\ref{eq:Wdiscrete}) can be written as 
\begin{align}
   \hat{\cal W}(\sqrt{\tau})
\approx &(\hat{I} \otimes {\rm e}^{-{\rm i}\hat{H}S_s}) \notag \\
 \times &
\prod_{l=-M_s}^{M_s} \left[\hat{\cal A}_l(\sqrt{\tau}) (\hat{I} \otimes {\rm e}^{{\rm i}\hat{H}\Delta_s})\right]
\prod_{l=M_s}^{-M_s} \left[(\hat{I} \otimes {\rm e}^{-{\rm i}\hat{H}\Delta_s})\hat{\cal A}_l(\sqrt{\tau}) \right]\notag\\
 \times& (\hat{I} \otimes {\rm e}^{ {\rm i}\hat{H}S_s}),
 \label{eq:Wdiscrete2}
\end{align}
where 
$
\hat{\cal A}_l(\sqrt{\tau})=
(\hat{R}_{Z_{\rm a}}(\varphi(s_l))\otimes I)
\times 
{\exp}\left[{-{\rm i}\frac{\sqrt{\tau}}{2} \Delta_s |f(s_l)| \hat{X}_{\rm a} \otimes \hat{A}}\right]\notag \\
\times 
(\hat{R}_{Z_{\rm a}}^\dagger(\varphi(s_l))\otimes I)$.
The expression in Eq.~(\ref{eq:Wdiscrete2}) makes it clear that the approximate $\hat{\cal W}(\sqrt{\tau})$ can be implemented by alternating between the system Hamiltonian time evolution with step size $\pm \Delta_s$ and the system-ancilla interaction $\hat{\cal A}_l(\sqrt{\tau})$. 

At the boundaries of the Trotterized sequence in Eq.~(\ref{eq:Wdiscrete}) (i.e., at $l=-M_s$), $\hat{\cal V}(s_{-M_s})$ and $\hat{\cal V}^\dagger(s_{-M_s})$ introduce coherent evolutions $\hat{I}\otimes{\rm e}^{{\rm i}\hat{H}S_s}$ and $\hat{I}\otimes{\rm e}^{-{\rm i}\hat{H}S_s}$ on the system, as appearing in the first and the third lines in Eq.~(\ref{eq:Wdiscrete2}). 
These operations cancel when $\hat{\cal W}(\sqrt{\tau})$ is applied repeatedly, except during the very first and last applications. 
%We remove these coherent evolutions because 
Since $\hat{I}\otimes{\rm e}^{{\rm i}\hat{H}S_s}$ merely alters the initial state, and $\hat{I}\otimes{\rm e}^{-{\rm i}\hat{H}S_s}$ produces only an overall phase on the ground state of $\hat{H}$, 
we remove these endpoint coherent evolutions from the first and last segments of the Trotterized implementation, following Ref.~\cite{Ding2024}. 
% Besides the dissipative channel, we also introduce coherent time evolution of the system, as described in Sec.~\ref{sec:discrete_time_evolution}.

\section{Quantum Experiments}\label{sec:results}

\subsection{Hamiltonian}

To demonstrate the protocol described above, we consider the 1D transverse-field Ising model, 
\begin{equation}
\hat{H}=J\sum_{i=0}^{N-2}\hat{Z}_{i}\hat{Z}_{i+1}
+B_{X}\sum_{i=0}^{N-1}\hat{X}_i,  
\label{eq:htfi}
\end{equation}
where $N$ is the number of system qubits, and $J$ and $B_X$ denote the exchange and transverse-field strengths, respectively. 
We study systems of sizes $N=4,\, 6$, and $19$ under open boundary conditions, using the parameters $J=-1$ and $B_X=-1.2$.

\subsection{Jump operator and filter function}

% Provided $\hat{A}$ in the jump operator is a product of Pauli operators, the unitary, ${\rm e}^{-{\rm i}\frac{\sqrt{\tau}}{2} \Delta_s |f(s_l)| \hat{X}_a \otimes \hat{A}}$, in Eq.~\eqref{eq:Usl} can be implemented with a phase gadget~\cite{Cowtan2020} up to single-qubit rotations. 
The choice of the jump operator plays a crucial role in determining the convergence behavior of the dissipative dynamics toward the steady state.
For certain spin and fermionic systems, it has been shown that bulk dissipation, in which the jump operators act on all system sites, drives the system to the ground state with a convergence time that scales logarithmically with the system size~\cite{zhan2025rapidquantumgroundstate}.
Importantly, when $\hat{A}$ is a product of $n_A$ Pauli operators acting on the system qubits, the unitary ${\rm e}^{-{\rm i}\frac{\sqrt{\tau}}{2} \Delta_s |f(s_l)| \hat{X}_{\rm a} \otimes \hat{A}}$ in Eq.~\eqref{eq:Usl} can be implemented using $2n_A$ CNOT gates along with appropriate single-qubit gates; this construction is referred to as a Pauli gadget~\cite{Cowtan2020}.
Here, for simplicity of implementation, we follow Ref.~\cite{Ding2024} and choose the jump operator to be $\hat{A}=\hat{Z}_0$, i.e., $n_A=1$ with the Pauli-$Z$ operator acting on the first qubit of the spin chain. 
With this choice, the interaction term between the ancilla and the system, ${\rm e}^{-{\rm i}\frac{\sqrt{\tau}}{2} \Delta_s |f(s_l)| \hat{X}_{\rm a} \otimes \hat{A}}$, can be implemented straightforwardly using elementary quantum gates.

As a filter function $\tilde{f}(\omega)$, we employ
\begin{equation}
\tilde{f}(\omega)= n_{\rm F}(\beta(\omega-b))-
 n_{\rm F}(\beta(\omega-a)), 
 \label{eq:filter}
\end{equation}
where $n_{\rm F}(x)=1/({\rm e}^{x}+1)$ is the Fermi-Dirac distribution function, 
%$ n_{\rm F}(\beta(\omega-a))=1/({\rm e}^{\beta(\omega-a)}+1)$,
and $a$, $b$, and $\beta$ are free parameters. 
Throughout this work, we assume $a<b<0$ and $\beta>0$.  
% It should be reminded that $\beta$ is merely a parameter of the filter function and nothing to do with temperatures. 
To ensure that the dissipative dynamics generated by the jump operator $\hat{K}$ prepares the ground state of the Hamiltonian, the filter function must have support only on the negative-frequency side [see Eq.~(\ref{eq:fwcond})], so that $\hat{K}$ suppresses transitions to higher-energy states. 
In the limit $\beta \to \infty$, the function in Eq.~(\ref{eq:filter}) approaches an ideal rectangular window and exactly satisfies Eq.~(\ref{eq:fwcond}). 
For finite $\beta$, however, the edges of the window are rounded: $\tilde{f}(\omega) \approx1$ for  $a+\frac{1}{\beta} \lesssim \omega \lesssim  b - \frac{1}{\beta}$ and $\tilde{f}(\omega) \approx 0$ for $\omega \lesssim a-\frac{1}{\beta}$ or  $b + \frac{1}{\beta} \lesssim  \omega$. 
We refer to this $O(1/\beta)$ rounding relative to the rectangular window as the genuine broadening of the filter function.

\begin{figure*}
\includegraphics[width=1.0\textwidth]{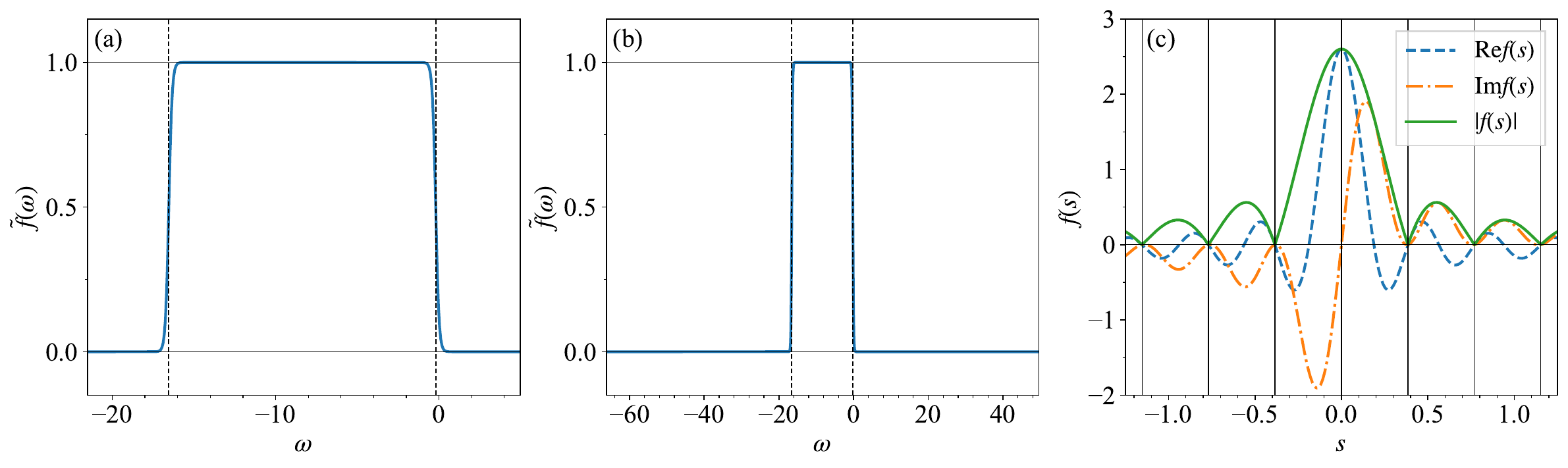}
\includegraphics[width=1.0\textwidth]{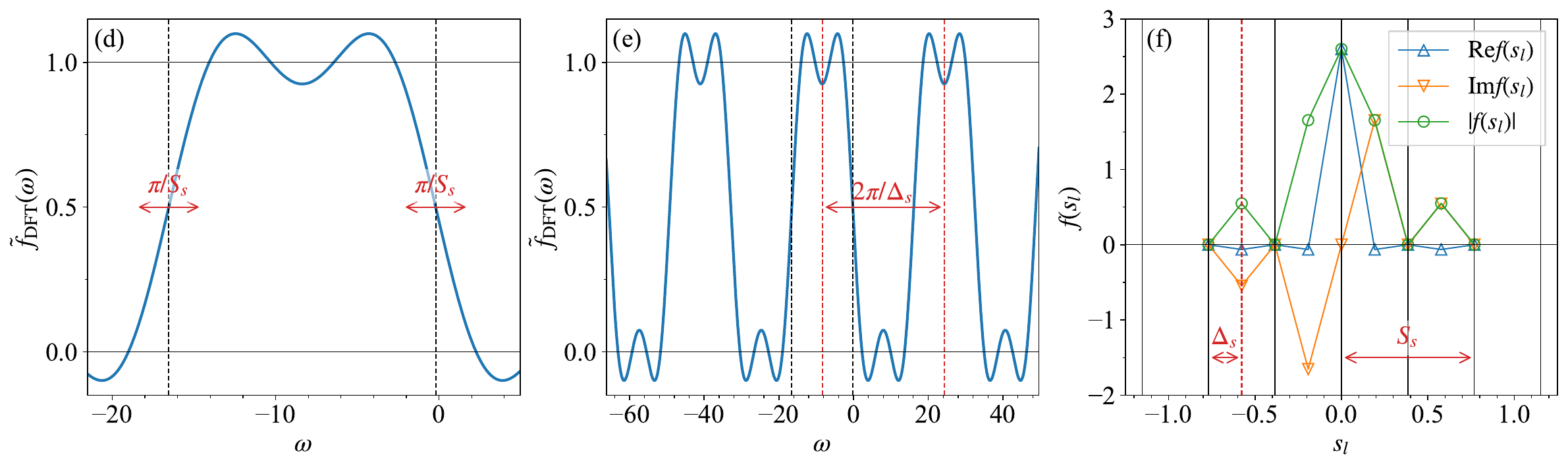}
\caption{
(a) Example of the filter function $\tilde{f}(\omega)$ in the frequency domain. 
The dashed vertical lines indicate $\omega=a$ and $\omega=b$ with $a <b<0$.
(b) Same as (a), but shown over a wider frequency range. 
(c) Fourier transform $f(s)$ of the filter function $\tilde{f}(\omega)$. 
The quantities $|f(s)|$, ${\rm Re}f(s)$, and ${\rm Im}f(s)$ are shown as solid, dashed, and dash-dotted curves, respectively. 
The solid vertical lines mark $s=2\pi n/(b-a)$ for integer $n$, where $f(s)=0$ except at $s=0$. 
(d-f) Same as (a-c), but with time discretization. 
In (d), the red arrows at $\omega=a$ and $\omega=b$ indicate the edge broadening of width $\pi/S_s$, arising from truncation of the time-integration range. 
In (e), the red arrow indicates the aliasing period $2\pi/\Delta_s$, corresponding to the discrete sampling points of $f(s_l)$ shown in (f). 
In (f), the large and small arrows indicate the time-discretization parameters $S_s$ and $\Delta_s$, respectively. 
The discretized values $f(s_l)$ shown in (f) are used in the quantum experiments for the $N=6$ system.
\label{fig:filters}}
\end{figure*}

The Fourier transform of the filter function is given by 
\begin{equation}
f(s)=
{\rm e}^{-{\rm i}\frac{b+a}{2}s}
\frac{\sin{\left(\frac{b-a}{2}s\right)}}{\beta\sinh{\left(\frac{\pi}{\beta}s\right)}}
\label{eq:filter_ft}
\end{equation}
with the normalization $f(0)=(b-a)/2\pi$ (see Appendix~\ref{app:derivation} for a detailed derivation). 
For finite values of $\beta$, the modulus $|f(s)|$ decays exponentially with increasing $|s|$ due to the hyperbolic sine function $\sinh{\left(\frac{\pi}{\beta}s\right)}$ in the denominator.
In contrast, the sine function in the numerator leads to zeros at $s=2\pi n/(b-a)$ for any nonzero integer $n$. 
As a guideline, the parameters $\beta$, $b$, and $a$ can be chosen as follows. 
The scale $1/\beta$ should be on the order of the excitation gap of the Hamiltonian; $b$ should be a negative value on the order of $1/\beta$; and $a$ should be a negative value on the order of the spectral radius of the Hamiltonian. 
In the quantum experiments presented in this work, we set $\beta=8/\Delta$, $b=-2/\beta=-\Delta/4$, and $a=-2|E_0|$, where $\Delta=E_{g_0}-E_0$ denotes the excitation gap. 
Because the system sizes considered here are sufficiently small, we evaluate $E_0$ and $E_{g_0}=E_1$ using exact diagonalization. 
Note that $|a|=2|E_0|$ corresponds to twice the spectral radius of $\hat{H}$. 
Numerical examples of the filter functions $\tilde{f}(\omega)$ and $f(s)$ for the $N=6$ system are shown in Figs.~\ref{fig:filters}(a) and \ref{fig:filters}(b), and Fig.~\ref{fig:filters}(c), respectively.

\subsection{Truncation and discretization of the OFT}
\label{sec:OFT}

As described in Sec.~\ref{sec:TD_OFT}, implementing $\hat{\cal W}(\sqrt{\tau})$ in Eq.~(\ref{eq:Wsa}) on a quantum computer requires discretizing the associated time integral as in Eqs.~(\ref{eq:calK_discrete}) and (\ref{eq:Wdiscrete}). 
We truncate the integration range using $S_s=4\pi/(b-a)$ and discretize it with $M_s=4$, giving $\Delta_s=S_s/M_s=\pi/(b-a)$. 
These values are chosen to ensure that the resulting circuit remains feasible within our available quantum computational resources. 
% , and not intended to estimate the ground state to a certain accuracy. 
The discrete Fourier transform (DFT) of $f(s)$,  
$\tilde{f}_{\rm DFT}(\omega)\equiv\sum_{l=-M_s}^{M_s}\Delta_s f(s_l){\rm e}^{{\rm i}\omega s_l}$, and the discretized values $f(s_l)$ for $|s_l|\leqslant S_s$ are shown in Figs.~\ref{fig:filters}(d) and \ref{fig:filters}(e), and Fig.~\ref{fig:filters}(f), respectively.  
Below, we describe how the DFT-based filter function $\tilde{f}_{\rm DFT}(\omega)$ 
%obtained via the DFT of the truncated and discretized $f(s_l)$ 
behaves as a consequence of truncating and discretizing $f(s)$. 
%However, it should be reminded that the function we use in quantum simulation is the time-discretized $f(s)$ shown in Fig.~\ref{fig:filters}(f). 

The truncation of the integration range ($S_s$)  in $f(s_l)$ induces oscillations in $\tilde{f}_{\rm DFT}(\omega)$ and effectively broadens the edges of $\tilde{f}_{\rm DFT}(\omega)$ at $\omega=a$ and $\omega=b$. 
A nonvanishing value of $\tilde{f}(\omega)$ at $\omega > 0$, i.e., a violation of Eq.~(\ref{eq:fwcond}), implies that the jump operator $\hat{K}$ contains energy-increasing transitions, in which case the ground state is no longer the steady state. This leads to a systematic error even in noiseless simulations. 
As shown in Fig.~\ref{fig:filters}(d), the broadening of the edges caused by the truncation can be estimated as $\sim \pi/S_s$, which is larger than the intrinsic broadening $1/\beta$ of the exact filter function in Eq.~\eqref{eq:filter}. 
This observation suggests choosing $S_s= O(\beta)$ in order to capture the genuine broadening of the filter function. Indeed, Appendix~\ref{app:leakage} shows that the truncation error decays as ${\rm e}^{-\pi S_s/\beta}$.

The discretization of the integral ($\Delta_s$) in $f(s)$ induces aliasing in $\tilde{f}_{\rm DFT}(\omega)$ with a period $\omega_{\rm alias}=2\pi/\Delta_s$. 
As observed in Fig.~\ref{fig:filters}(e), the separation between the right edge of the filter function around the support of the original filter and the left edge of the spurious right-adjacent copy is approximately $\omega_{\rm alias}-(b-a)$, where we ignore the broadening of the edges.
For the present choice of $\Delta_s$, we have $\omega_{\rm alias}=2(b-a)$, and thus the separation is about $b-a$, which is nearly twice the spectral radius of $\hat{H}$.
This implies that the error due to discretization should be insignificant.
As shown in Appendix~\ref{app:alias}, the discretization error decays as ${\rm e}^{-\beta(\omega_{\rm alias}-2\|\hat{H}\|)}$.

\subsection{Discrete time evolution} 
\label{sec:discrete_time_evolution}

We consider a single step of the discrete-time evolution, 
\begin{equation}
\hat{\rho}(m+1) = {\Gamma}_H\circ{\Gamma}_{K}\circ{\Gamma}_H[\hat{\rho}(m)], 
\end{equation}
where
\begin{equation}
    {\Gamma}_{H}[\hat{\rho}] = 
    [{\rm e}^{-{\rm i}\hat{H}\Delta_t}]^{N_t/2}
    \hat{\rho} 
    [{\rm e}^{{\rm i}\hat{H}\Delta_t}]^{N_t/2}.
\end{equation}
implements a coherent unitary time evolution under the transverse-field Ising Hamiltonian for a duration $t/2=N_t\Delta_t/2$.
% with the initial state for the system qubits denoted by  $\hat{\rho}(0)$.
We interleave dissipative and coherent evolution in order to accelerate convergence toward the ground state, as observed in Refs.~\cite{Ding2024, Brunner2025, Fang2025}.
The short-time propagator ${\rm e}^{-{\rm i}\hat{H}\Delta_t}$ for a small time step $\Delta_t$ is approximated using the second-order Trotter formula,
${\rm e}^{-{\rm i}\hat{H}\Delta_t}=
{\rm e}^{-{\rm i}\hat{H}_X\Delta_t/2}
{\rm e}^{-{\rm i}\hat{H}_{ZZ}\Delta_t}
{\rm e}^{-{\rm i}\hat{H}_X\Delta_t/2}
+O(\Delta_t^3)
$, 
where $\hat{H}_{\rm ZZ}$ and $\hat{H}_X$ denote the Ising and transverse-field terms of the Hamiltonian $\hat{H}$ in Eq.~(\ref{eq:htfi}), respectively. 
An analogous Trotterization is applied to the short-time evolution ${\rm e}^{{\rm i}\hat{H}\Delta_s}$ that appears in $\Gamma_K$
[see Eqs.~(\ref{eq:sigmatau}), (\ref{eq:KrausCS}), (\ref{eq:Wdiscrete}), and (\ref{eq:VdagV})].

\subsection{Experimental details}

We use Quantinuum's trapped-ion quantum computer \Reimei{}~\cite{reimei_benchmark} to evaluate the energy expectation value along the dissipative evolution~\footnote{The experiments were conducted in June 2025 and November 2025.}.
At the time of the experiments, the \Reimei{} system consisted of $20$ qubits and natively supported single-qubit rotation gates and a parametrized two-qubit gate $R_{ZZ}(\alpha) = e^{-\frac{1}{2} {\rm i} \pi \alpha \hat{Z}_i \hat{Z}_j}$, where $\alpha$ is a real-valued parameter. 
The native two-qubit gate could be applied to arbitrary pairs of qubits. 
The average infidelities of the single-qubit and two-qubit gates were approximately $0.004\%$ and $0.14\%$, respectively, and the average state-preparation-and-measurement (SPAM) error was around $0.35\%$. 
All quantum circuits used in the experiments were compiled using \texttt{TKET}~\cite{Sivarajah2020}.

\subsection{Energy expectation values}

\begin{figure}
    \centering
    \includegraphics[width=0.48\textwidth]{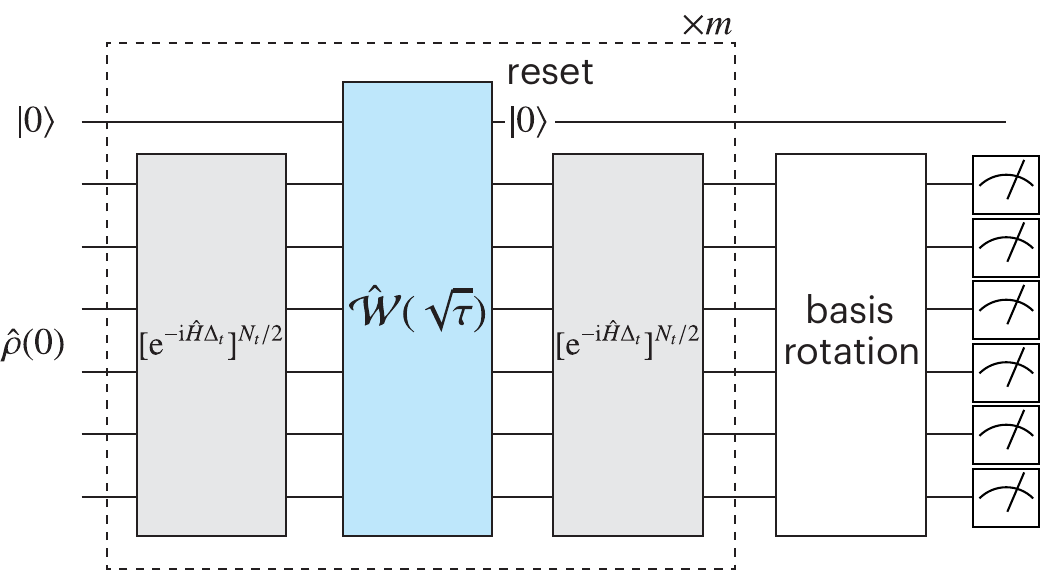}
    \caption{Quantum circuit used to estimate the expectation value ${\rm Tr}[\hat{\rho}(m)\hat{O}]$. 
    The topmost qubit, initialized in the state $|0\rangle$, serves as the ancilla, while the remaining qubits, initialized in the state $\hat{\rho}(0)$, serves as the system qubits.
    }
    \label{fig:circuit}
\end{figure}

Figure~\ref{fig:circuit} shows the quantum circuit used to estimate the expectation value ${\rm Tr}[\hat{\rho}(m)\hat{O}]$ of an observable $\hat{O}$. 
The first qubit serves as the ancilla, and the remaining six qubits represent the system. 
The initial state is prepared as $|0\rangle\langle0|_{\rm a}\otimes\hat{\rho}(0)$. 
The gray box denotes the coherent unitary evolution of the system under the Hamiltonian $\hat{H}$, 
while the blue box represents the dilated unitary $\hat{\cal W}(\sqrt{\tau})$. 
The reset of the ancilla qubit, which discards information about the measurement outcome, effectively implements the partial trace over the ancilla, provided that a sufficiently large number of shots is taken. 
The white box corresponds to a basis rotation to measure the observable $\hat{O}$, after which the system qubits are measured in the computational basis. 
For example, the basis rotation is the identity when measuring $\hat{O}=\hat{Z}_i\hat{Z}_{i+1}$, whereas it consists of Hadamard gates when measuring $\hat{O}=\hat{X}_i$. 
The total number of SPAM operations in the circuit, including qubit initializations and resets, is $N_{\rm SPAM}=m+2N+1$, where $m$ accounts for the resets of the ancilla qubit, $N+1$ corresponds to the initializations, and $N$ arises from the final measurements of the system qubits.

\begin{figure}
\includegraphics[width=0.5\textwidth]{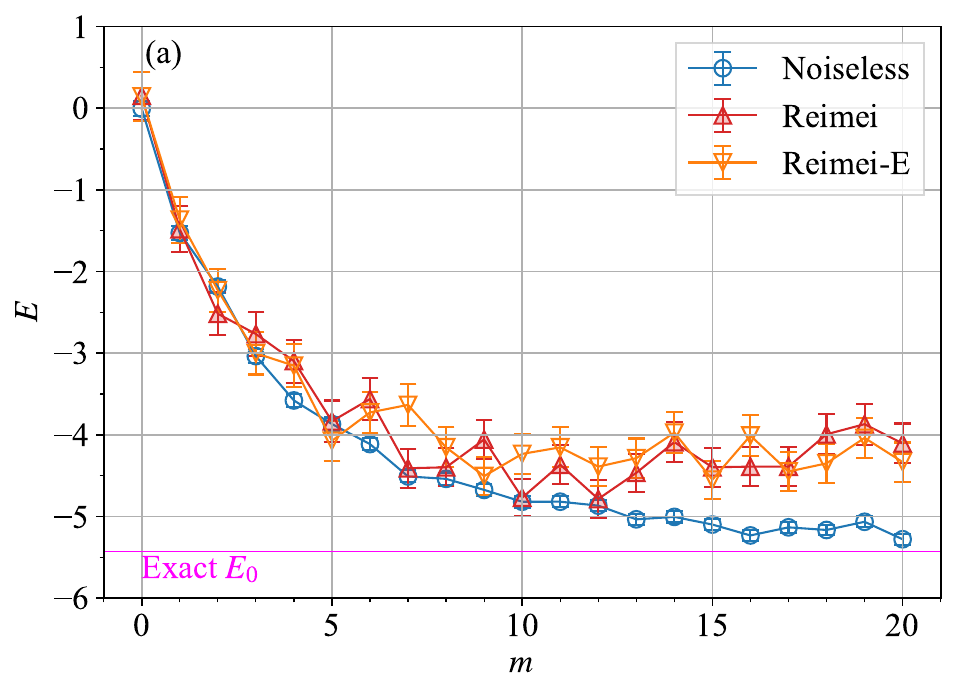}
\includegraphics[width=0.5\textwidth]{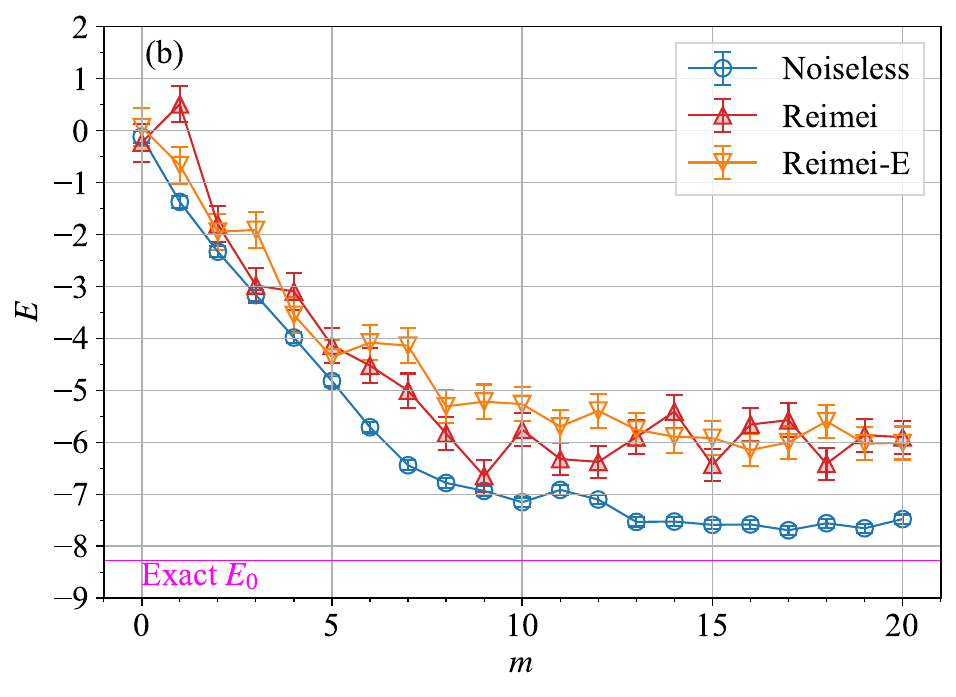}
\caption{Energy $E(m)$ as a function of the time steps $m$ for (a) $N=4$ and (b) $N=6$. The parameters $J=-1$ and $B_X=-1.2$ are used for the transverse-field Ising model in both cases. 
Blue circles, 
red filled triangles, and 
orange inverted triangles show results from 
noiseless simulations (Noiseless), 
the quantum hardware (\Reimei{}), and 
the noisy emulator (Reimei-E),  
respectively. 
For each data point, noiseless results are obtained using 1000 measurement shots, whereas \Reimei{} and Reimei-E results are obtained using 100 measurement shots. 
Error bars indicate the standard deviations.
The exact ground-state energy $E_0$ is shown as a magenta horizontal line. 
\label{fig:energy}}
\end{figure}

Figure~\ref{fig:energy} shows the energy expectation value 
\begin{equation}
    E(m) = {\rm Tr}\left[\hat{\rho}(m)\hat{H}\right]
\end{equation}
of the transverse-field Ising model as a function of the time step $m$. 
We set the parameters of the dynamics to $\tau=4$, $\Delta_t=0.25$, and $N_t=4$. 
The choice $\tau=4$ is motivated by noiseless simulations for the $N=6$ system, where it yields the fastest convergence to the steady state among the values of $\tau$ that we examined (see Appendix~\ref{app:taudep}). 
The initial state is prepared as the product state of Pauli-$Y$ eigenstates, $\hat{\rho}(0)=\prod_{i}[(\hat{I}_i+\hat{Y}_i)/2]$. 
Since the Hamiltonian contains only Pauli-$X$ and $Z$ operators, the corresponding energy expectation value at $m=0$ is $E(0)={\rm Tr}\left[\hat{\rho}(0)\hat{H}\right]=0$.
Experiments are performed on \Reimei{}, and noiseless simulations are carried out using the Aer simulator of \texttt{Qiskit}~\cite{qiskit}. 
We also include results obtained from the noisy \Reimei{} emulator (denoted as Reimei-E), which closely reproduces the behavior of the hardware. 
The number of measurement shots is set to $N_{\rm shots}=100$ for \Reimei{} and Reimei-E, and $N_{\rm shots}=1000$ for the noiseless Aer simulator. 
Using a single ancilla qubit, the total number of qubits required in the experiments is $N+1$.

First, the results obtained from Reimei-E agree well with the hardware results of \Reimei{}, remaining within a few standard deviations for both $N=4$ and $N=6$, as shown in Fig.~\ref{fig:energy}. 
The number of native two-qubit gates (i.e., $R_{ZZ}$ gates) scales as $N_{\rm 2Q}=57m$ for $N=4$ and $N_{\rm 2Q}=79m$ for $N=6$. Thus, the largest circuits at $m=20$ contain $1140$ and $1580$ $R_{ZZ}$ gates for $N=4$ and $N=6$, respectively. 
Given the two-qubit gate fidelity of $1-1.4\times10^{-3}=0.9986$, a naive estimate of the overall circuit fidelity at $m=20$ would be $(0.9986)^{1140}\approx0.2$ for $N=4$ and $(0.9986)^{1580}\approx0.11$ for $N=6$.  
Because $\hat{H}$ is traceless, the energy expectation value of the completely mixed state $\hat{I}/d$ is ${\rm Tr}[\hat{H}]/d=0$. 
If we assume that the noisy state at $m=20$ is described by a mixture of the ground state and the completely mixed state---with the former contributing a fraction of 0.2 or 0.11 (as in the depolarizing channel)---then we would expect $E(m=20)={\rm Tr}\left[\hat{\rho}(20)\hat{H}\right]\sim 0.2\times E_0$ or $0.11\times E_0$, where $E_0$ denotes the exact ground-state energy. 
However, the experimental signals in Fig.~\ref{fig:energy} are significantly better than these naive estimates. 
As discussed above, the dissipative channel $\Gamma_K$ has the ground state as the steady state of the dynamical map, which suggests an intrinsic robustness of the protocol to noise. 
Indeed, we observe that a steady state persists even in the presence of non-negligible hardware noise. 
Although noise may shift the steady state from the ideal noiseless one $\hat{\rho}^*$ to another higher-energy state $\hat{\rho}^*_{\rm noisy}$, this noisy state is not the completely mixed state $\hat{\rho}^*_{\rm noisy}\not = \hat{I}/d$, 
as evidenced by Fig.~\ref{fig:energy}, where the energy expectation values converge to finite values rather than zero. 
Thus, a reasonably good approximation of the ground state can still be prepared despite the presence of noise. 
A related analysis of robustness can be found in Ref.~\cite{Brunner2025}. 
We also note that even the noiseless simulation does not converge exactly to the true ground-state energy due to systematic discretization errors.

\subsection{Zero-noise extrapolation}
We now assess the ability of ZNE to mitigate errors. 
We adopt the gate-folding method applied to the native two-qubit gates, i.e., the $R_{ZZ}(\theta)$ gates. 
In this method, each $R_{ZZ}(\theta)$ gate in a compiled circuit is replaced by 
\begin{equation}
R_{ZZ}(\theta)  \mapsto  
\left[R_{ZZ}(\theta) R_{ZZ}(-\theta)\right]^{\frac{G-1}{2}} R_{ZZ}(\theta), 
\label{eq:zne}
\end{equation}
where the noise-scaling factor $G$ is an odd integer. 
Specifically, we consider circuits with $G=1$, $3$, and $5$, where $G=1$ corresponds to the original circuit. 
We first compile the $G=1$ circuit for the target device using the Pytket SDK~\cite{Sivarajah2020} with the most aggressive optimization setting \verb|optimisation_level=3|. 
For $G=3$ and $5$, we then replace each $R_{ZZ}$ gate in the compiled circuit according to Eq.~(\ref{eq:zne}).
For extrapolation, we employ both linear and exponential fits of the expectation values as a function of $G$. 
The fitting functions take the form $\tilde{a}G+\tilde{b}$ (linear) and $\tilde{a} \exp(\tilde{b}G)$ (exponential), where $\tilde{a}$ and $\tilde{b}$ are fitting parameters. 
We emphasize that the exponential fit~\cite{Endo2018} is an appropriate choice, as the circuit fidelity is expected to decrease exponentially with $G$. 
The linear fit is expected to be consistent with the exponential fit in the small-noise regime.
Energy expectation values are then estimated by extrapolating the data at two points ($G=1$ and $3$) using the linear fit, or at three points ($G=1$, $3$, and $5$) using the exponential fit, to the zero-noise limit $G\to 0$.

\begin{figure}
\includegraphics[width=0.5\textwidth]{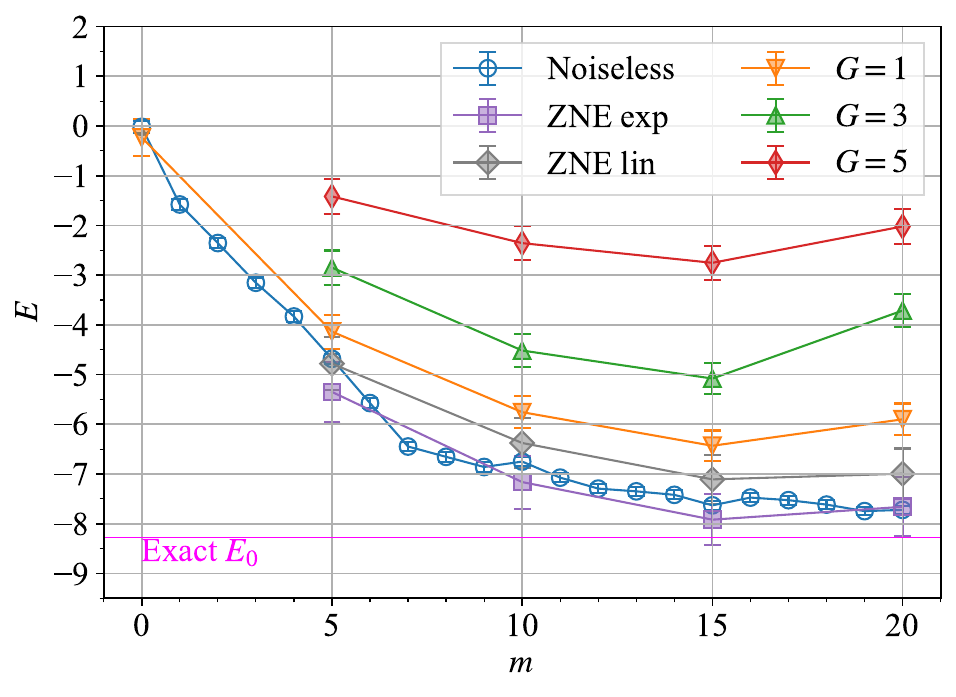}
\caption{Energy $E(m)$ as a function of the time step $m$ for the $N=6$ transverse-field Ising model with parameters $J=-1$ and $B_X=-1.2$. 
Results for noise-scaling factors $G=1$, $3$, and $5$ are obtained on \Reimei{}.
For each data point, noiseless results use $1000$ measurement shots, whereas the \Reimei{} results for $G=1$, $3$, and $5$ use $100$ measurement shots. 
``ZNE exp" and ``ZNE lin" denote the ZNE estimates obtained via exponential and linear extrapolations, respectively. 
Error bars indicate the standard deviations and, for the ZNE exp and ZNE lin data, also include the uncertainty associated with the extrapolation. 
The exact ground-state energy $E_0$ is indicated by the magenta horizontal line. 
\label{fig:energyZNE}}
\end{figure}

Figure~\ref{fig:energyZNE} shows the ZNE results for the $N=6$ system, using the same model parameters as in Fig.~\ref{fig:energy}. 
Noisy data for $G=1$, $3$, and $5$ are obtained on \Reimei{}, with $N_{\rm shots}=100$ for each value of $m$ and $G$.  
We note that the results for $G=1$ in Fig.~\ref{fig:energyZNE} are the same as those shown in Fig.~\ref{fig:energy}(b). 
The number of $R_{ZZ}$ gates in the circuit at $m$th time step scales as $N_{\rm 2Q}=79Gm$. 
At $m=0$, ZNE data are not shown because $N_{\rm 2Q}(m=0)=0$. 
For $m \geqslant 15$, ZNE with the exponential fit exhibits better agreement with the noiseless results than the linear fit, although in most cases the two extrapolation schemes agree within statistical uncertainties. 
At the largest time step $m=20$, the relative error of the noisy $G=1$ data, defined as $|1-E_{G=1}(m)/E_{\rm Noiseless}(m)|$, is approximately $0.3$, where the subscripts (``$G=1$" and ``Noiseless") match the legend labels used in Fig.~\ref{fig:energyZNE}. 
This error is successfully eliminated by the exponential ZNE extrapolation. 
Details of the fitting procedures are provided in Appendix~\ref{app:zne}.

\begin{figure}
\includegraphics[width=0.5\textwidth]{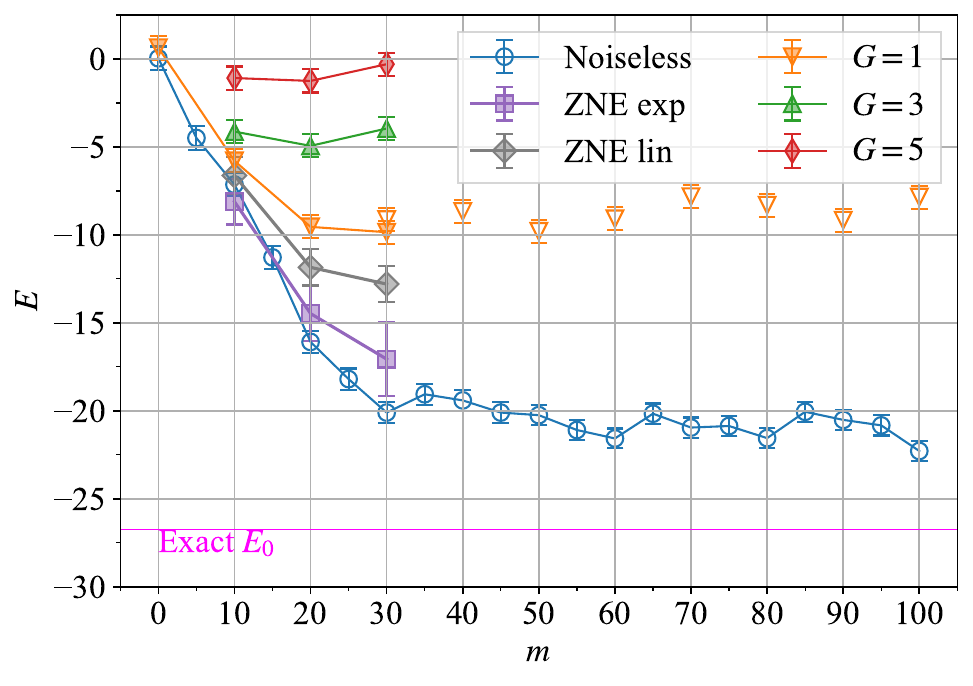}
\caption{Energy $E(m)$ as a function of the time step $m$ for the $N=19$ transverse-field Ising model with parameters $J=-1$ and $B_X=-1.2$. 
Results for noise-scaling factors $G=1$, $3$, and $5$ for $m\leqslant 30$ are obtained on \Reimei{}. 
For comparison, results obtained with Reimei-E for $G=1$ at $m \geqslant 30$ are also shown as orange open triangles.  
For each data point, noiseless results use $1000$ measurement shots, whereas the \Reimei{} results for $G=1$, $3$, and $5$ and the Reimei-E results for $G=1$ at $m\geqslant 30$ use $100$ measurement shots. 
``ZNE exp" and ``ZNE lin" denote the ZNE estimates obtained via exponential and linear extrapolations, respectively.
Error bars indicate the standard deviations and, for the ZNE exp and ZNE lin data, also include the uncertainty associated with the extrapolation. 
The exact ground-state energy $E_0$ is indicated by the magenta horizontal line. 
\label{fig:energyZNE19}}
\end{figure}

Figure~\ref{fig:energyZNE19} shows the ZNE results for the $N=19$ system, using the same model parameters as in Fig.~\ref{fig:energyZNE}. 
Noisy data for $G=1$, $3$, and $5$ are obtained on \Reimei{} for time steps up to $m\leqslant 30$, with $N_{\rm shots}=100$ for each value of $m$ and $G$. 
The number of $R_{ZZ}$ gates in the circuit at time step $m$ scales as $N_{\rm 2Q}=137Gm$, and thus the largest circuits at $m=30$ contain $4110$, 12330, and 20550 $R_{ZZ}$ gates for $G=1, 3,$ and $5$, respectively. 
Although the discrepancy between the exact ground-state energy $E_0$ and the noiseless simulation results is more pronounced than in the $N=6$ case, the overall behavior remains qualitatively similar.
The larger discrepancy arises from the truncation of the time integral at $S_s$, which leads to an $O(1/S_s)$ broadening of the filter-function edges. 
The ZNE extrapolation using the exponential fit shows consistently better agreement with the noiseless results than the linear fit, particularly at $m=20$ and $30$. 
Even at $m=30$, the ZNE estimate agrees with the noiseless value within two standard deviations, although the agreement is less evident than at earlier times ($m=10$ and $20$). 
We note that the present ZNE can mitigate only noise associated with two-qubit gate errors. 
Other errors that cannot be mitigated by ZNE, such as state-preparation and measurement errors, memory errors, and leakage errors, may become non-negligible at $m=30$ compared with $m=10$ and $20$.
At $m=30$, the relative error in the noisy $G=1$ data, quantified as $|1-E_{G=1}(m)/E_{\rm Noiseless}(m)|~\sim 0.5$, is reduced by approximately a factor of three when applying ZNE with the exponential fit: $|1-E_{{\rm ZNE\ exp}}(m)/E_{\rm Noiseless}(m)|~\sim 0.15$. 
Details of the fitting procedure are provided in Appendix~\ref{app:zne}.

Due to the limitation on the maximum number of quantum gate operations allowed per circuit on \Reimei{} and Reimei-E, we were unable to perform ZNE---particularly for $G=3$ and $5$---for time steps $m\geqslant 40$. 
Instead, we carry out noisy simulations using Reimei-E with $G=1$ for $m\geqslant 30$. 
The results are shown as orange open triangles in Fig.~\ref{fig:energyZNE19}. 
Remarkably, even at $m=100$, where the circuit contains $N_{\rm 2Q}=13 700$ $R_{ZZ}$ gates, we still observe a nonzero energy expectation value. 
This further demonstrates the robustness of the dissipative ground-state preparation protocol against noise.

\section{Summary and Discussion}~\label{sec:discussion}

We have demonstrated a dissipative ground-state preparation protocol for a one-dimensional transverse-field Ising model on the trapped-ion quantum computer \Reimei{}. 
To this end, we derived a Kraus representation of the corresponding dissipation channel and showed that the fidelity with respect to the ground state is monotonically nondecreasing under successive applications of the channel to any initial state.

Specifically, we applied the protocol to estimate the ground-state energy of the transverse-field Ising model for system sizes up to $N=19$ sites. 
To implement the protocol, we used $N+1$ qubits of the trapped-ion quantum computer, where the additional qubit serves as the ancilla required to realize the dissipative dynamics via the Stinespring dilation unitary. 
In the presence of hardware noise, the measured energy expectation values did not converge to those obtained from noiseless simulations. 
To address this deviation, we employed ZNE and obtained energy estimates that agree with the noiseless results within statistical uncertainties. 
We found that ZNE with the exponential extrapolation yields better agreement with the noiseless results than linear extrapolation. 
For the $N=6$ system, the errors are almost completely eliminated by ZNE, whereas for the $N=19$ system, the errors at the largest time step are reduced by a factor of approximately three.

We determined the parameters $\beta$, $b$, and $a$ for the filter function using the exact diagonalization method. 
Since the one-dimensional transverse-field Ising model can be solved efficiently, these parameters can readily be obtained even for lager system sizes. 
For situations where exact solutions are not available (e.g., mixed-field Ising models), quantum Monte Carlo or tensor-network methods may be used to estimate $\beta$, $b$, and $a$. 
In addition, for certain parameter regimes of the Hubbard and Heisenberg models, values of $E_0$ may be found in the literature~\cite{LeBlanc2015,Zheng2017,Nomura2017,Ido2018,Seki2019,Sorella2023,Sandvik1997,Iqbal2013,Iqbal2016,Nomura2021}, 
which can also be used to estimate the spectral radius. 
Importantly, the key quantities to estimate are the spectral radius and the energy gap. The parameters $\beta^{-1}$ and $b$ should be on the order of the energy gap, while $b-a$ should be on the order of the spectral radius. 
Overestimating (underestimating) the spectral radius leads to a larger circuit depth (a larger aliasing error) than necessary, whereas overestimating (underestimating) the energy gap results in a larger leakage error (a larger circuit depth) than necessary.
We emphasize, however, that precise estimates of these low-lying eigenenergies are not strictly required, as long as $\beta$, $a$, and $b$ are chosen to (approximately) satisfy the condition on the filter function in Eq.~(\ref{eq:fwcond}). 
It is also worth noting that, in the NISQ era with limited quantum resources, the modification of the filter function arising from time discretization can have a more significant impact on the systematic error of the protocol than modest inaccuracies in estimating these parameters.

While we have fixed the form of the filter function and the time-discretization scheme as in Eq.~(\ref{eq:calK_discrete}), these choices may also be optimized to achieve fast convergence while minimizing quantum resources, such as the number of two-qubit gates. 
For instance, one may adopt a coarser time-discretization grid at early time steps $m$ to reduce resource overhead, and then refine the grid at later time steps to more accurately reproduce the step-function-like behavior of $\tilde{f}(\omega)$. 
Such an optimization is conceptually analogous to optimized imaginary-time evolution in classical computing~\cite{Beach2019,Sorella2023} and represents a promising direction for future developments in dissipative ground-state preparation.

\begin{acknowledgments}
We thank Anshuman Bhardwaj, Haruki Emori,  Masanao Ozawa, and Maho Nakata for valuable discussions.
We also thank Clemens Gneiting for helpful comments.
K.~S. is supported by JSPS KAKENHI Grants No. JP22K03520 and No. JP26K06954.
T.~H. is supported by JSPS KAKENHI Grants No. JP24K00630 and No. JP25K01002. 
S.~Y. is supported by JSPS KAKENHI Grant No. JP24K02948.
This work is based in part on results obtained from a project (JPNP20017), subsidized by the New Energy and Industrial Technology Development Organization (NEDO), Japan. 
We also acknowledge funding from JST COI-NEXT (Grant No. JPMJPF2221).
Furthermore, we acknowledge support from the UTokyo Quantum Initiative, the RIKEN TRIP initiative (RIKEN Quantum), and the COE research grant in computational science from Hyogo Prefecture and Kobe City through the Foundation for Computational Science.

\end{acknowledgments} 

 \section*{Data Availability}
 The experimental and numerical data that support the findings of this study are available at Zenodo~\cite{seki_2026_18334125}.

\appendix

\section{Derivation of the Kraus representation} 
\label{app:derivation_Kraus}

In this Appendix, we derive the Kraus representation of $\Gamma_K$ given on the right-hand side of Eq.~(\ref{eq:KrausCS}) and provide additional remarks on its structure.

\subsection{Derivation of Eq.~(\ref{eq:KrausCS})}

We begin by considering the Taylor expansion of the dilated unitary operator $\hat{\cal W}(\sqrt{\tau})$ defined in Eq.~(\ref{eq:W}):
\begin{equation}
\hat{\cal W}(\sqrt{\tau})
= \sum_{k=0}^{\infty} c_k\hat{\cal K}^k
= \sum_{k=0}^{\infty} \left(c_{2k}\hat{\cal K}^{2k} +c_{2k+1}\hat{\cal K}^{2k+1}\right),
\label{eq:WTaylor}
\end{equation}
where $c_k=(-{\rm i}\sqrt{\tau})^{k}/k!$. 
The decomposition into even and odd powers is made for clarity. 
From Eq.~(\ref{eq:calK}), we obtain 
\begin{align}
\hat{\cal K}^{2k} &= 
|0\rangle\langle0|_{\rm a}\otimes \left(\hat{K}^\dagger \hat{K}\right)^k +
|1\rangle\langle1|_{\rm a}\otimes \left(\hat{K}\hat{K}^\dagger\right)^k, 
\label{eq:Keven} \\
\hat{\cal K}^{2k+1}  &= 
|1\rangle \langle 0|_{\rm a} \otimes \hat{K} \left(\hat{K}^\dagger \hat{K}\right)^k +  
|0\rangle \langle 1|_{\rm a} \otimes \left(\hat{K}^\dagger \hat{K}\right)^k \hat{K}^\dagger ,
  \label{eq:Kodd}
\end{align}
showing that $\hat{\cal K}^{2k}$ is diagonal and $\hat{\cal K}^{2k+1}$ is off-diagonal in the ancilla-qubit subspace. 
Thus, the matrix representation of $\hat{\cal W}(\sqrt{\tau})$ reads 
\begin{align}
    \hat{\cal W}(\sqrt{\tau}) 
& = \begin{bmatrix}
 \sum_{k=0}^{\infty}c_{2k}(\hat{K}^\dagger\hat{K})^k & \sum_{k=0}^{\infty}c_{2k+1}(\hat{K}^\dagger\hat{K})^k\hat{K}^\dagger \\
 \hat{K} \sum_{k=0}^{\infty}c_{2k+1}(\hat{K}^\dagger\hat{K})^k& 
 \sum_{k=0}^{\infty}c_{2k}(\hat{K}\hat{K}^\dagger)^k 
 \end{bmatrix}
 \notag \\
 & = \begin{bmatrix}
 \hat{C} & 
 -{\rm i}\sqrt{\tau}\hat{S}_{\rm c}\hat{K^\dagger}\\
 -{\rm i}\sqrt{\tau}\hat{K}\hat{S}_{\rm c}& 
 \hat{C}' 
 \end{bmatrix} \notag \\
  & = \begin{bmatrix}
 \hat{M}_0 & 
 -\hat{M}_1^\dagger\\
 \hat{M}_1& 
 \hat{C}' 
 \end{bmatrix}, 
\end{align}
where the cosine and cardinal-sine expansions defined in Eqs.~(\ref{eq:Cos}) and (\ref{eq:Sinc}) have been used, and $\hat{C}'=\cos\sqrt{\tau \hat{K}\hat{K}^\dagger}$. 
In the final expression, we identify the Kraus operators  
$\hat{M}_{0}=\langle 0|\hat{\cal W}(\sqrt{\tau})|0\rangle_{\rm a}=\hat{C}$ and 
$\hat{M}_{1}=\langle 1|\hat{\cal W}(\sqrt{\tau})|0\rangle_{\rm a}=-{\rm i}\sqrt{\tau}\hat{K}\hat{S}_{\rm c}$.

Using the block-matrix form of $\hat{\cal W}(\sqrt{\tau})$, Eq.~(\ref{eq:sigmatau}) can be written as 
\begin{align}
\hat{\sigma}(\tau)
&
=\hat{\cal W}(\sqrt{\tau})
\left(|0\rangle\langle0|_{\rm a}\otimes \hat{\rho}\right)
\hat{\cal W}^\dagger(\sqrt{\tau}) \notag \\
&
= \begin{bmatrix}
 \hat{M}_0 & 
 -\hat{M}_1^\dagger\\
 \hat{M}_1& 
 \hat{C}' 
 \end{bmatrix}
 \begin{bmatrix}
 \hat{\rho} & 
 0\\
 0& 
 0 
 \end{bmatrix}
 \begin{bmatrix}
 \hat{M}_0^\dagger & 
 \hat{M}_1^\dagger\\
 -\hat{M}_1& 
 \hat{C}'^\dagger 
 \end{bmatrix}
 \notag \\
&
= \begin{bmatrix}
 \hat{M}_0\hat{\rho}\hat{M}_0^\dagger & 
 \hat{M}_0\hat{\rho}\hat{M}_1^\dagger\\
 \hat{M}_1\hat{\rho}\hat{M}_0^\dagger& 
 \hat{M}_1\hat{\rho}\hat{M}_1^\dagger
 \end{bmatrix}.
 \label{eq:sigmatau_app}
\end{align}
Tracing out the ancilla then yields Eq.~(\ref{eq:KrausCS}): 
$\Gamma_K[\hat{\rho}]=
{\rm Tr}_{\rm a}\left[\hat\sigma(\tau)\right]=\hat{M}_0\hat{\rho}\hat{M}_0^\dagger+\hat{M}_1\hat{\rho}\hat{M}_1^\dagger$, 
which is the Kraus representation of the map.

\subsection{Lindblad dynamics}

As a remark on $\Gamma_K$, we note that its lowest-order expansion coincides with a Lindblad dynamics with the jump operator $\hat{K}$. 
By explicitly extracting the terms proportional to $\tau^0$ and $\tau^1$ from Eq.~(\ref{eq:KrausCS}), we obtain
\begin{equation}
\Gamma_K[\hat{\rho}]
= \hat{\rho}+\tau{\hat{K}}\hat{\rho}\hat{K}^\dagger - \frac{\tau}{2}\left\{\hat{K}^\dagger\hat{K},\hat{\rho} \right\} + O(\tau^2).
\label{eq:Lindblad}
\end{equation}
Therefore, in the small $\tau$ limit, $\Gamma_K$ reduces to the Lindblad evolution generated by the jump operator $\hat{K}$, as discussed in Ref.~\cite{Ding2024}.

\section{Fourier transform of the filter function}
\label{app:derivation}

In this Appendix, we derive the Fourier transform $f(s)$ of the filter function $\tilde{f}(\omega)$, following the definitions in Eqs.~(\ref{eq:fs}) and (\ref{eq:filter}). This derivation leads directly to the expression given in Eq.~(\ref{eq:filter_ft}). 
%\subsection{$s=0$}

First, at $s=0$, we obtain 
\begin{align}
f(0)=
\int_{-\infty}^{\infty}
\frac{{\rm d}\omega}{2\pi} 
\left[
n_{\rm F}(\beta(\omega-b))-
n_{\rm F}(\beta(\omega-a))
\right] 
=
\frac{b-a}{2\pi},
\label{eq:f0}
\end{align}
which is independent of $\beta$. 
In deriving Eq.~(\ref{eq:f0}), we have used the identity $n_{\rm F}(\beta(\omega-b))=\partial_\omega \left[-\frac{1}{\beta}\ln(1+{\rm e}^{-\beta(\omega-b)})\right]$.

\begin{figure}
\includegraphics[width=0.4\textwidth]{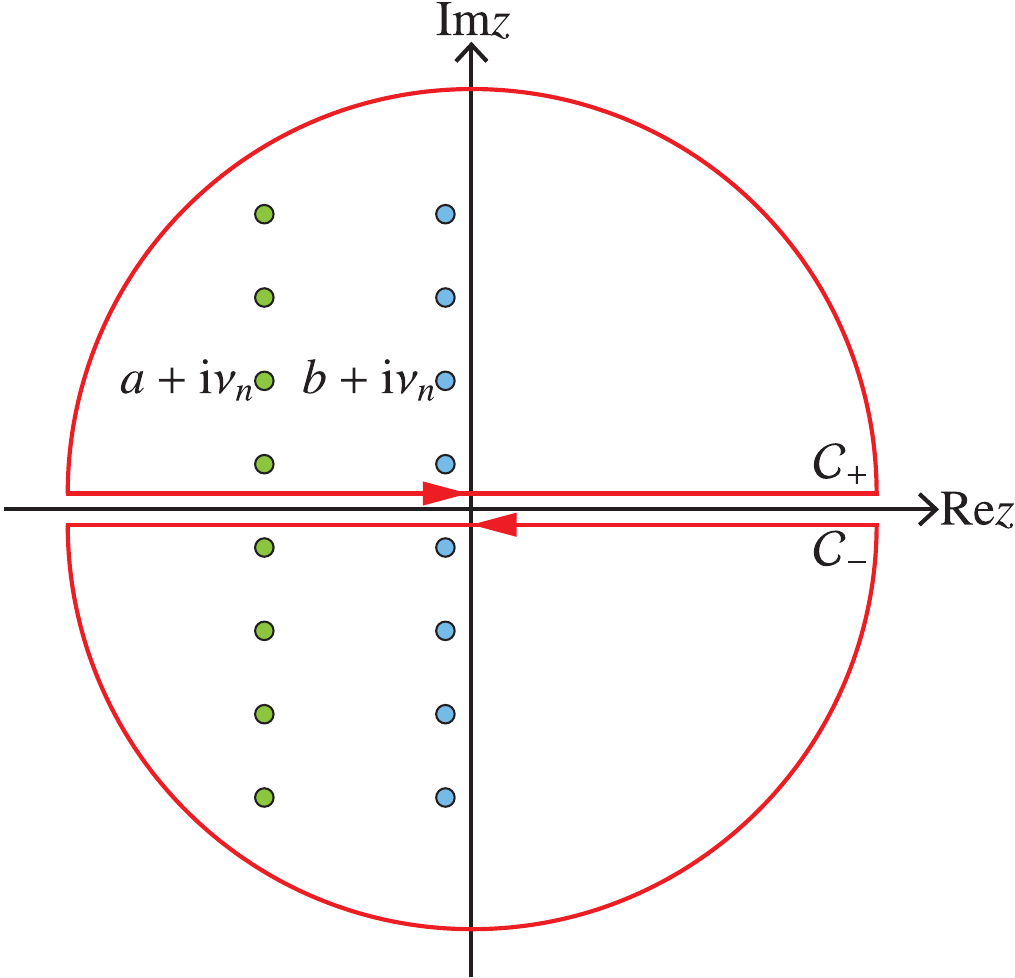}
\caption{Schematic illustration of the contours $C_+$ and $C_-$ in the complex $z$-plane used to evaluate the Fourier transform $f(s)$ of the filter function. The green and blue dots denote the shifted fermionic Matsubara frequencies $a+{\rm i}\nu_n$ and $b+{\rm i}\nu_n$, respectively.
\label{fig:contour}}
\end{figure}

Next, to evaluate $f(s)$ for $s\not=0$, we employ contour integration in the complex $z$-plane and use the fact that $n_{\rm F}(\beta(z-a))$ has poles at 
\begin{equation}
z = a + {\rm i} \nu_n, \qquad {\rm i} \nu_n= {\rm i}\frac{(2n+1)\pi {\rm }}{\beta},
\end{equation}
each with residue $-1/\beta$, where $\nu_n$ are the fermionic Matsubara frequencies and $n$ is an integer~\cite{Ezawa1957}. 
Depending on the sign of $s$, we rewrite $f(s)$ as a contour integral: 
\begin{equation}
f(s) = 
\begin{dcases}
\frac{1}{2\pi}\oint_{{\cal C}_{+
}} {\rm d} z \tilde{f}(z) {\rm e}^{-{{\rm i}zs}} \quad (s<0),\\
-\frac{1}{2\pi}\oint_{{\cal C}_{-}} {\rm d} z \tilde{f}(z) {\rm e}^{-{{\rm i}zs}} \quad (s>0),
\end{dcases}
\label{eq:fs_contours}
\end{equation}
where the contour ${\cal C}_{+}\ ({\cal C}_{-})$ consists of the real axis together with an infinite semicircular arc enclosing the upper (lower) half of the complex $z$-plane in the counterclockwise direction (see Fig.~\ref{fig:contour}). 
The radius of the semicircle is taken to infinity at the end of the calculation. 
The choice of contour is dictated by the factor ${\rm e}^{-{\rm i}zs}$: for $s<0$ ($s>0)$, the contribution from the arc at infinity vanishes only when the contour is closed in the upper (lower) half-plane. 
Since ${\cal C}_{+}$ (${\cal C}_{-}$) encloses the poles at $a+{\rm i}\nu_n$ and $b+{\rm i}\nu_n$ in the upper (lower) half-plane, 
the residue theorem gives 
\begin{equation}
f(s) = 
\begin{dcases}
-\frac{\rm i}{\beta}
\left(
{\rm e}^{-{{\rm i}bs}}-{\rm e}^{-{{\rm i}as}}
\right)
\sum_{n=0}^{\infty} {\rm e}^{\frac{(2n+1)\pi}{\beta} s} \quad (s<0),\\
\frac{\rm i}{\beta}
\left(
{\rm e}^{-{{\rm i}bs}}-{\rm e}^{-{{\rm i}as}}
\right)
\sum_{n=0}^{\infty}{\rm e}^{-\frac{(2n+1)\pi}{\beta} s}  \quad (s>0).
\end{dcases}
\label{eq:fscases}
\end{equation}
To unify these expressions, we use the geometric-series representation of the hyperbolic cosecant, 
\begin{equation}
\frac{1}{\sinh{s}} 
= \frac{2}{{\rm e}^{s}-{\rm e}^{-s}} 
=
\begin{dcases}
-2\sum_{n=0}^{\infty} {\rm e}^{(2n+1)s} \quad (s<0),\\
2\sum_{n=0}^{\infty}{\rm e}^{-(2n+1) s}  \quad (s>0),
\end{dcases}
\label{eq:csch}
\end{equation}
which allows Eq.~(\ref{eq:fscases}) to be written compactly as 
\begin{equation}
f(s)=\frac{{\rm i}(
{\rm e}^{-{{\rm i}bs}}-{\rm e}^{-{{\rm i}as}}
)}{2\beta\sinh{\frac{\pi}{\beta}s}}
=
{\rm e}^{-{\rm i}\frac{b+a}{2}s}
\frac{\sin{\frac{b-a}{2}s}}{\beta\sinh{\frac{\pi}{\beta}s}}.
\end{equation}
Finally, we readily verify that $\lim_{s\to0} f(s)=(b-a)/2\pi$, which exactly matches the value of $f(0)$ obtained in Eq.~(\ref{eq:f0}), confirming that $f(s)$ is continuous at $s=0$. 
Thus we obtain Eq.~(\ref{eq:filter_ft}).

\section{Time discretization of the OFT}

We analyze the errors arising from the regularization of the OFT in Eq.~\eqref{eq:Kt}, which include the time-discretization (aliasing) error, $\epsilon_{\rm alias}$, and the truncation (leakage) error, $\epsilon_{\rm leak}$.

\subsection{Time discretization}
\label{app:alias}

We begin by analyzing the time-discretization error, i.e., the aliasing error. 
To this end, we employ Poisson's resummation formula,
\begin{align}
\begin{split}
    &\Delta_s\sum_{l=-\infty}^{\infty}f(l\Delta_s)\hat{A}(l\Delta_s)
    \\
    &=
    \int_{-\infty}^{\infty}{\rm d}t f(t)\hat{A}(t)
    +
    \sum_{k\in\mathbb{Z}\backslash\{0\}}\int_{-\infty}^{\infty}{\rm d}t\,
    {\rm e}^{-{\rm i}2\pi kt/\Delta_s} f(t)\hat{A}(t),
\end{split}
\end{align}
which leads to the aliasing error 
\begin{align}
\begin{split}
    \epsilon_{\rm alias}
    &=\bigg\|
        \int_{-\infty}^{\infty}{\rm d}t f(t)\hat{A}(t)
        -
        \Delta_s\sum_{l=-\infty}^{\infty}f(l\Delta_s)\hat{A}(l\Delta_s)
    \bigg\|
    \\
    &=
    \bigg\|
        \sum_{k\in\mathbb{Z}\backslash\{0\}}\int_{-\infty}^{\infty}{\rm d}t\,
        {\rm e}^{-{\rm i}2\pi kt/\Delta_s} f(t)\hat{A}(t)
    \bigg\|
    \\
    &=
    \bigg\|
        \sum_{\nu\in B_H}\sum_{k\in\mathbb{Z}\backslash\{0\}}\int_{-\infty}^{\infty}{\rm d}t\,
        {\rm e}^{-{\rm i}2\pi kt/\Delta_s} f(t){\rm e}^{{\rm i}\nu t}\hat{A}_\nu
    \bigg\|
    \\
    &=
    \bigg\|
        \sum_{\nu\in B_H}\sum_{k\in\mathbb{Z}\backslash\{0\}}\tilde{f}(\nu-2\pi k/\Delta_s)\hat{A}_\nu
    \bigg\|.
\end{split}
\end{align}
Here, $\|\cdot\|$ denotes the spectral norm, $B_H$ is the set of all eigenvalue differences of $\hat{H}$, and $\hat{A}_\nu\equiv \sum_{i,j:\, E_i-E_j=\nu}|E_i\rangle\langle E_i|\hat{A}|E_j\rangle\langle E_j|$.
For the filter function defined in Eq.~\eqref{eq:filter}, 
we obtain the bound 
\begin{align}
\begin{split}
    \epsilon_{\rm alias}
    &\leqslant
    4^N\|\hat{A}\| 
        \sum_{k\in\mathbb{Z}\backslash\{0\}}\tilde{f}(2\|\hat{H}\|-2\pi k/\Delta_s).
\end{split}
\end{align}
The sum evaluates to 
\begin{align}
\begin{split}
    &\sum_{k\in\mathbb{Z}\backslash\{0\}}\tilde{f}(2\|\hat{H}\|-2\pi k/\Delta_s)
    \\
    &=
    \sum_{k\in\mathbb{Z}\backslash\{0\}}
    \frac{1}{{\rm e}^{-\beta(2\pi k/\Delta_s-2\|\hat{H}\|+b)}+1} 
    - 
    \frac{1}{{\rm e}^{-\beta(2\pi k/\Delta_s-2\|\hat{H}\|+a)}+1}
    \\
    &\leqslant
    {\rm e}^{2\beta\|\hat{H}\|} ({\rm e}^{-\beta a}- {\rm e}^{-\beta b})
    \sum_{k>0}{\rm e}^{-2\pi \beta k/\Delta_s} 
    \\
    &\quad+
    {\rm e}^{-2\beta\|\hat{H}\|} ({\rm e}^{\beta a}- {\rm e}^{\beta b})
    \sum_{k<0}{\rm e}^{2\pi \beta k/\Delta_s} 
    \\
    &\leqslant
    {\rm e}^{2\beta\|\hat{H}\|} ({\rm e}^{-\beta a}- {\rm e}^{-\beta b})
    \frac{{\rm e}^{-2\pi\beta/\Delta_s}}{1-{\rm e}^{-2\pi\beta/\Delta_s}}.
\end{split}
\end{align}
Therefore, the aliasing error is bounded as
\begin{align}
    \epsilon_{\rm alias}
    \leqslant 
    \frac{\|\hat{A}\| ({\rm e}^{-\beta a}- {\rm e}^{-\beta b})}{1-{\rm e}^{-2\pi\beta/\Delta_s}}4^N{\rm e}^{-2\beta(\pi/\Delta_s-\|\hat{H}\|)}.
\end{align}

\subsection{Truncation of the infinite sum}
\label{app:leakage}

We now analyze the error arising from truncating the infinite sum, i.e., the leakage error: 
\begin{align}
\label{eq:error_leakage}
\begin{split}
    \epsilon_{\rm leak}
    &=
    \Big\|\sum_{|l|>M_s}f(l\Delta_s)\Big\|
    \leqslant
    \frac{4}{\beta\bigl(1-{\rm e}^{-\frac{2\pi\Delta_s}{\beta}M_s}\bigr)}
    \sum_{l=M_s}^{\infty}{\rm e}^{-\frac{\pi\Delta_s}{\beta}l}
    \\
    &=%\le
    \frac{4}{\beta\bigl(1-{\rm e}^{-\frac{2\pi}{\beta}S_s}\bigr) \bigl(1-{\rm e}^{-\frac{\pi\Delta_s}{\beta}}\bigr)}
    {\rm e}^{-\frac{\pi}{\beta}S_s},
\end{split}
\end{align}
where in the first inequality we used the bound $|f(s)|\leqslant 2\beta^{-1}\,{\rm e}^{-\frac{\pi}{\beta}s}(1-{\rm e}^{-\frac{2\pi}{\beta}S_s})^{-1}$ for $s\ge S_s$.

\section{$\tau$ dependence of the convergence}
\label{app:taudep}

\begin{figure}
\includegraphics[width=0.5\textwidth]{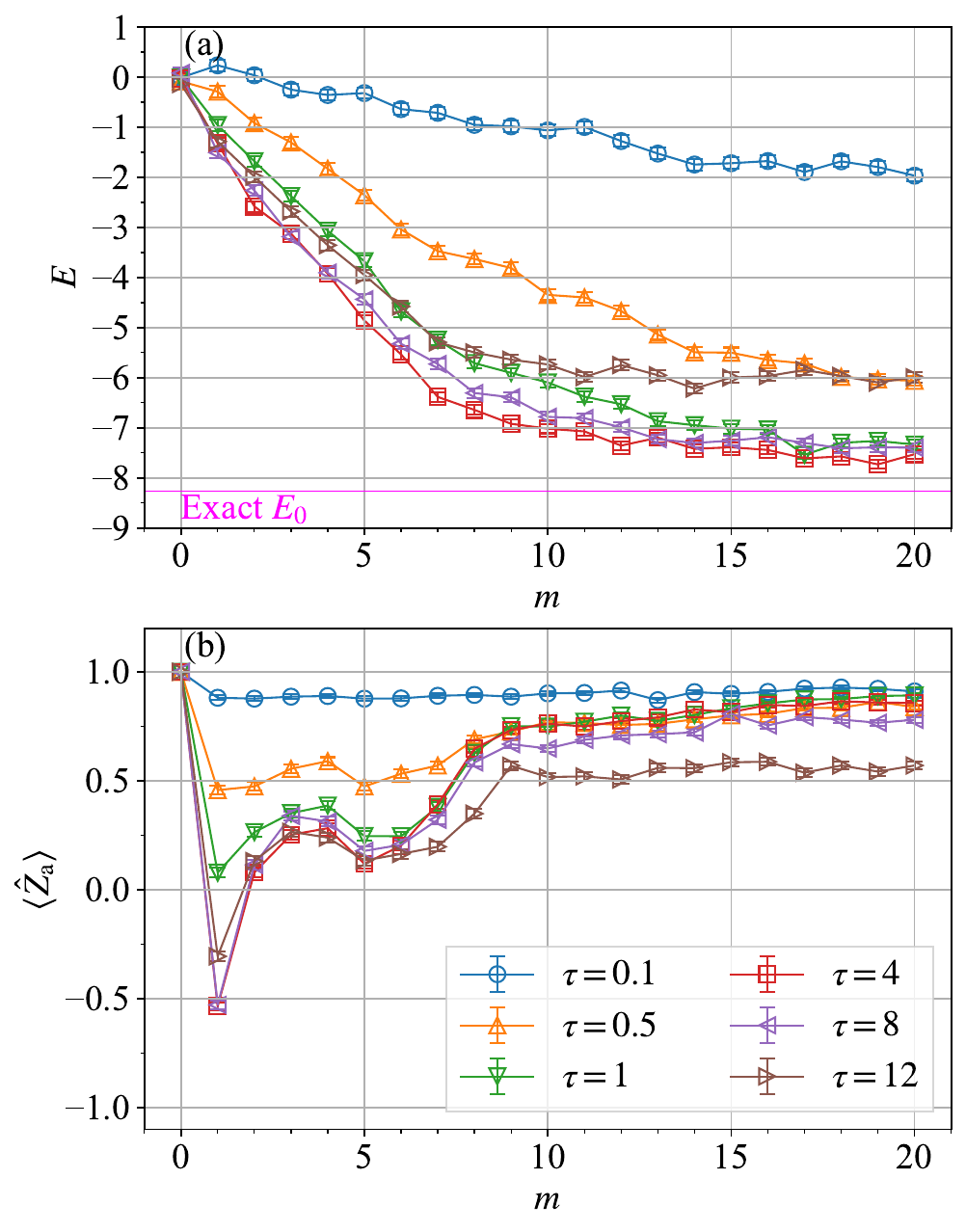}
\caption{Noiseless simulation results for (a) the energy $E(m)$ and (b) the expectation value of the ancilla Pauli-$Z$ operator as functions of the time steps $m$ for the $N=6$ transverse-field Ising model at various values of the dissipation time step~$\tau$. The model parameters are $J=-1$ and $B_X=-1.2$. 
\label{fig:energy_taudep}}
\end{figure}

In this Appendix, we examine how the convergence behavior depends on the parameter $\tau$, based on noiseless simulations. 
We recall that $\sqrt{\tau}$ plays the role of an ``interaction time" between the system and the ancilla, as seen in Eq.~(\ref{eq:Usl}). 
In practice, since the number of gates and resets increases proportionally with the time step $m$, it is important to choose a value of $\tau$ that achieves the fastest convergence as a function of $m$.

Figure~\ref{fig:energy_taudep}(a) shows the noiseless simulation results for the energy expectation value of the $N=6$ system for $\tau=0.1,0.5,1,4,8,$ and $12$, with the model parameters fixed to $J=-1$ and $B_X=-1.2$. 
We observe that the convergence rate with respect to the time step $m$ depends strongly on $\tau$. 
For small $\tau$ (e.g., $\tau=0.1$), the energy decreases only slowly as a function of $m$, whereas the convergence becomes progressively faster as $\tau$ increases. 
The most rapid decrease is obtained at $\tau=4$, and essentially the same behavior is seen at $\tau=8$. 
However, further increasing $\tau$ to $\tau=12$ results in slower convergence compared with the optimal region around $\tau=4$-$8$.

To gain further insight into the $\tau$-dependence of the convergence, we examine the probabilities of measuring the ancilla qubit in the states $|0\rangle_{\rm a}$ and $|1\rangle_{\rm a}$.
Equation~(\ref{eq:sigmatau_app}) implies that $\Gamma_K$ can be written as 
\begin{align}
\Gamma_K[\hat{\rho}]
&={\rm Tr}_{\rm a} 
\left[
|0\rangle\langle0|_{\rm a}\otimes \hat{M}_0\hat{\rho}\hat{M}_0^\dagger
+
|1\rangle\langle1|_{\rm a}\otimes \hat{M}_1\hat{\rho}\hat{M}_1^\dagger
\right],
\end{align}
where the off-diagonal terms in the ancilla subspace have been omitted, as they vanish upon tracing out the ancilla. 
The probabilities of obtaining outcomes $0$ and $1$ when measuring the ancilla qubit are therefore given by
\begin{align}
p_0&={\rm Tr}[ \hat{M}_0\hat{\rho}\hat{M}_0^\dagger],\\
p_1&={\rm Tr}[ \hat{M}_1\hat{\rho}\hat{M}_1^\dagger],
\end{align}
respectively.
The completeness relation in Eq.~(\ref{eq:Completeness}) ensures that $p_0+p_1=1$ for any normalized state satisfying ${\rm Tr}[\hat\rho]=1$.

Figure~\ref{fig:energy_taudep}(b) shows the noiseless simulation results for the expectation value $\langle \hat{Z}_{\rm a}\rangle=p_0-p_1=2p_0-1$ of the ancilla qubit at time step $m$. 
Here, $\langle \hat{Z}_{\rm a}\rangle$ is estimated from the measurement outcome of the ancilla qubit just before reset (classically controlled Pauli-$X$) operation at time step $m$. By definition, $\langle \hat{Z}_{\rm a}\rangle=1$ at $m=0$, since the ancilla is initialized in the state $|0\rangle_{\rm a}$. 
Note also that, once the system has converged to the ground state, the ancilla will be measure in $|0\rangle_{\rm a}$. 
At small $\tau$ (e.g., $\tau=0.1$), $\langle \hat{Z}_{\rm a}\rangle$ remains close to 1, indicating that the ancilla is almost always measured in the state $|0\rangle_{\rm a}$. 
As $\tau$ increases, $\langle \hat{Z}_{\rm a}\rangle$ decreases, reflecting an increased likelihood of obtaining the outcome $|1\rangle_{\rm a}$. 
In particular, the value of $\langle \hat{Z}_{\rm a} \rangle$ at the first time step $m=1$ exhibits the strongest dependence on $\tau$.  
Assuming that $\tau$ is small, the Kraus operators may be approximated as $\hat{M}_0\approx1-\frac{1}{2}\tau\hat{K}^\dagger \hat{K}$ and $\hat{M}_1\approx-{\rm i}\sqrt{\tau}\hat{K}$, using Eqs.~(\ref{eq:Cos}) and (\ref{eq:Sinc}). 
Accordingly, the ancilla outcome probabilities are approximated up to $O(\tau)$ as 
\begin{align}
p_0 &={\rm Tr}[ \hat{M}_0\hat{\rho}\hat{M}_0^\dagger] \approx 1 - \tau{\rm Tr}[\hat{K}\hat{\rho}\hat{K}^\dagger],\\
p_1 &={\rm Tr}[ \hat{M}_1\hat{\rho}\hat{M}_1^\dagger] \approx \tau {\rm Tr}[\hat{K}\hat{\rho}\hat{K}^\dagger].
\end{align}
These expressions are consistent with the numerical observation that $p_0$ decreases with increasing $\tau$ in the small-$\tau$ regime. 
The post-measurement states $\hat{\rho}_0$ and $\hat{\rho}_1$ of the system corresponding to ancilla outcomes $|0\rangle\langle0|_{\rm a}$ and $|1\rangle\langle1|_{\rm a}$ are given by 
\begin{align}
p_0\hat{\rho}_0 &=\hat{M}_0\hat{\rho}\hat{M}_0^\dagger \approx \hat{\rho} - \frac{\tau}{2}\left\{\hat{K}^\dagger\hat{K},\hat{\rho}\right\},\\
p_1\hat{\rho}_1 &=\hat{M}_1\hat{\rho}\hat{M}_1^\dagger \approx  \tau\hat{K}\hat{\rho}\hat{K}^\dagger,
\end{align}
which give the explicit dependence on $\tau$ to leading order. 
Although heuristic, the noiseless simulations suggest that increasing $\tau$ does increase $p_1$, i.e., the population of the branch associated with $\hat{\rho}_1$. However, a larger $p_1$ does not necessarily imply faster convergence of the overall dissipative dynamics. 
For general $\tau$, the Kraus operators involve trigonometric functions of $\sqrt{\tau \hat{K}^\dagger \hat{K}}$, and thus both the energy $E(m)$ and the ancilla expectation value $\langle \hat{Z}_{\rm a} \rangle$ can exhibit nonmonotonic dependence on $\tau$. 
Indeed, such nonmonotonic behavior is observed numerically when $\tau$ is increased up to $\tau=12$. 
This indicates the presence of an optimal $\tau$ that yields the fastest convergence to the steady state when the other parameters are fixed.

\begin{figure}
\includegraphics[width=0.5\textwidth]{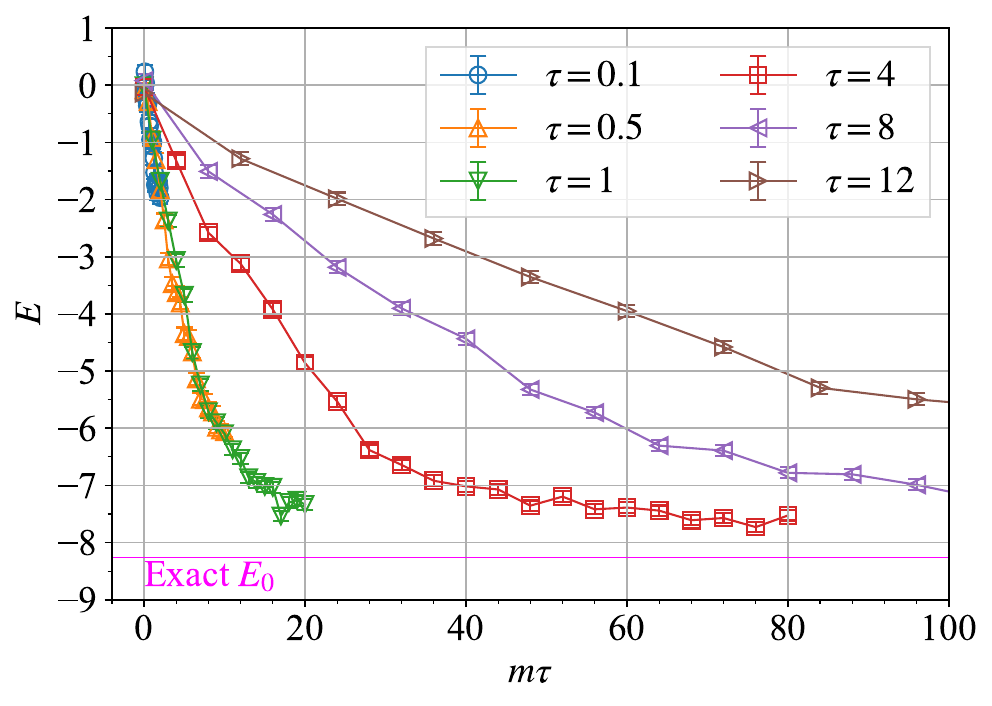}
\caption{
Same as Fig.~\ref{fig:energy_taudep}(a), except that the horizontal axis is $m\tau$. 
\label{fig:energy_mtaudep}}
\end{figure}

Finally, we examine whether the dissipative dynamics at a given $\tau$ lie within the Lindblad regime. 
Figure~\ref{fig:energy_mtaudep} displays the energy as a function of $m\tau$. 
If $\tau$ is sufficiently small for the Lindblad dynamics in Eq.~(\ref{eq:Lindblad}) to be valid, $m$ applications of the dissipation channel, $\Gamma_K^m(\tau)$, should coincide with a single application over a time interval $m\tau$, i.e., $\Gamma_K(m\tau)$. 
This implies that when an observable such as the energy is plotted as a function of $m\tau$ for different $\tau$, the data should collapse onto a single curve. 
The results for $\tau=0.1$, $0.5$, and $1$ indeed lie approximately on the same curve, indicating that these dynamics are within the Lindblad regime. 
By contrast, the results for $\tau=4$, $8$, and $12$ clearly deviate from those for smaller $\tau$, indicating that they lie beyond the Lindblad regime.

\section{Zero-noise extrapolation}
\label{app:zne}

\begin{figure*}
\includegraphics[width=0.495\textwidth]{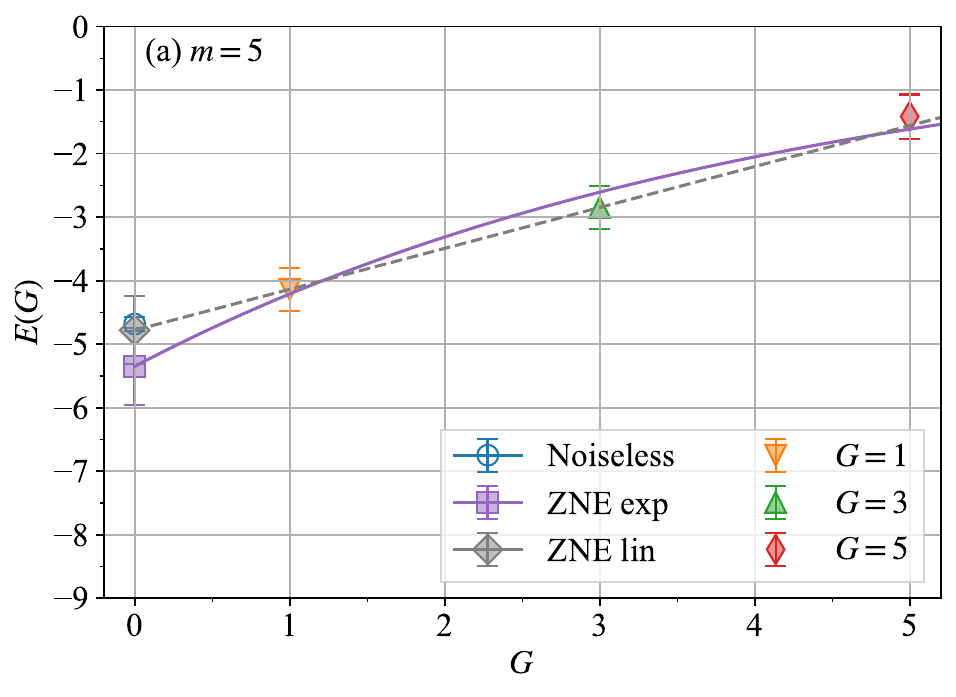}
\includegraphics[width=0.495\textwidth]{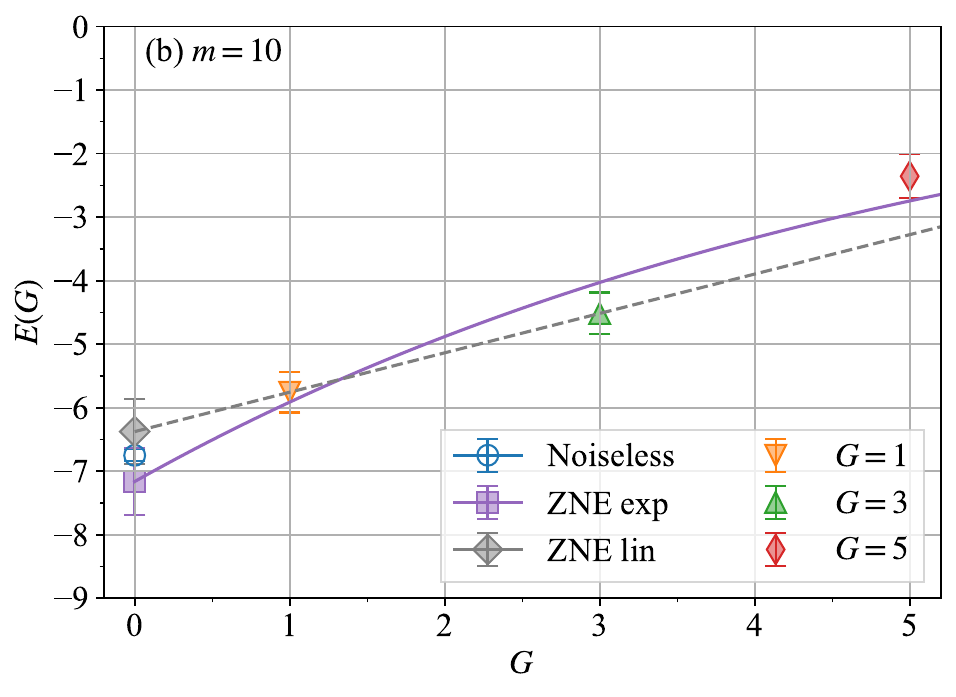}
\includegraphics[width=0.495\textwidth]{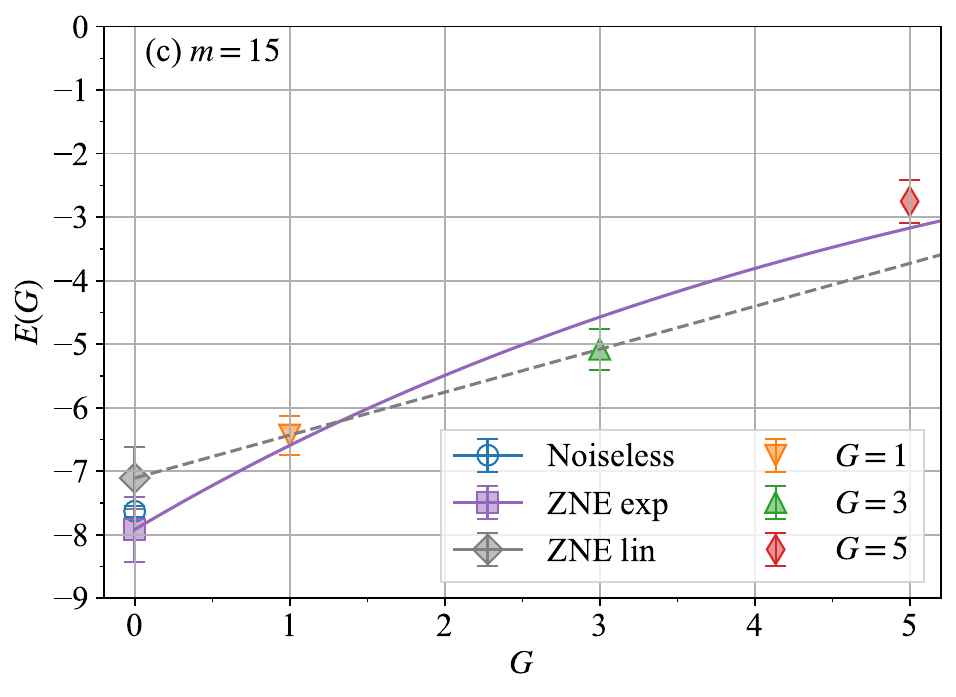}
\includegraphics[width=0.495\textwidth]{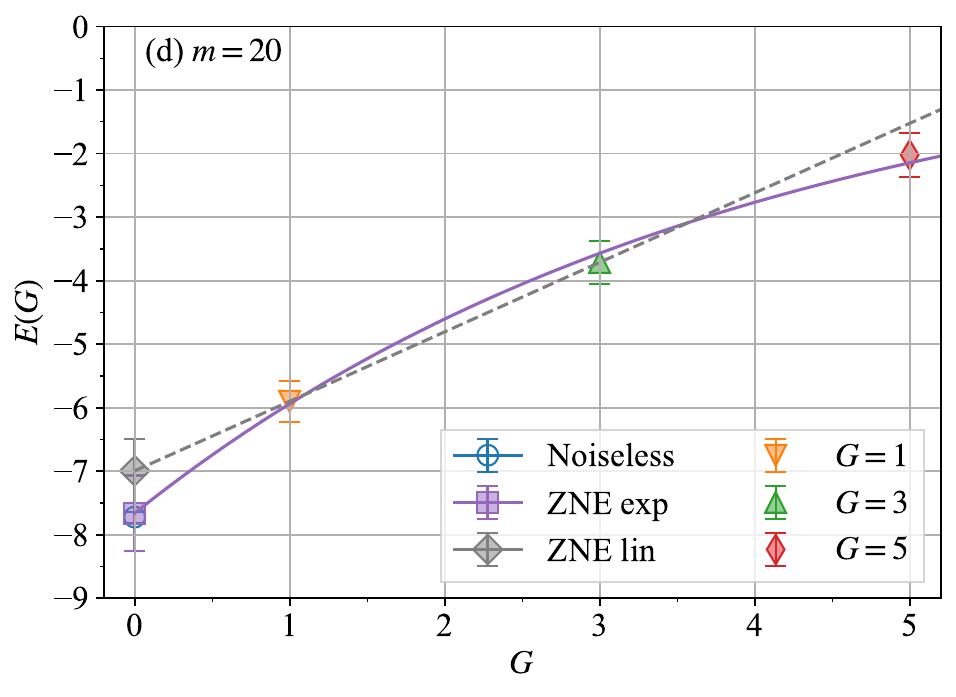}
\caption{Energy as a function of the noise-scaling factor $G$ at time steps (a) $m=5$,  (b) $m=10$, (c) $m=15$, and (d) $m=20$ for the $N=6$ transverse-field Ising model with parameters $J=-1$ and $B_X=-1.2$. 
Solid curves and dashed lines indicate the exponential fits, $\tilde{a} \exp({\tilde{b}G})$, and the linear fits, $\tilde{a}G + \tilde{b}$, respectively, applied to the experimental data. 
The data at $G=1$, $3$, and $5$ are used for the exponential fits, whereas only the data at $G=1$ and $3$ are used for the linear fits. 
Error bars at $G=1$, 3, and 5 indicate the standard deviations of the measurements, while those at $G=0$ (the extrapolated values) include the uncertainties arising from the extrapolation. 
Noiseless simulation results are also shown as blue circles at $G=0$. 
Each data point at $G=1$, $3$, and $5$ is obtained using 100 measurement shots, whereas the noiseless results are obtained using 1000 measurement shots. 
\label{fig:ZNEdetails}}
\end{figure*}

\begin{figure*}
\includegraphics[width=0.495\textwidth]{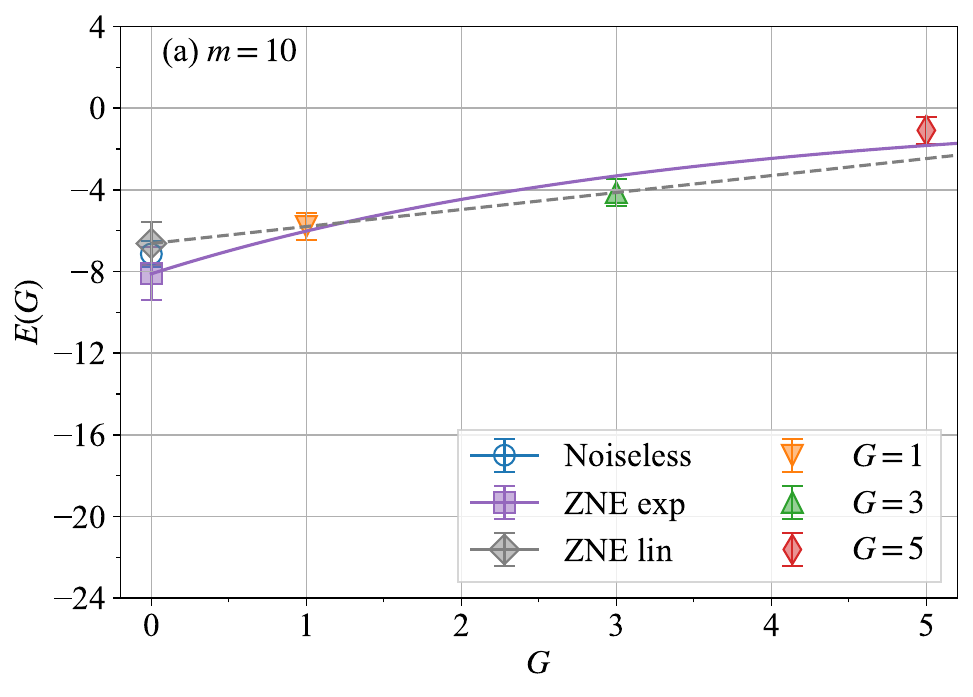}
\includegraphics[width=0.495\textwidth]{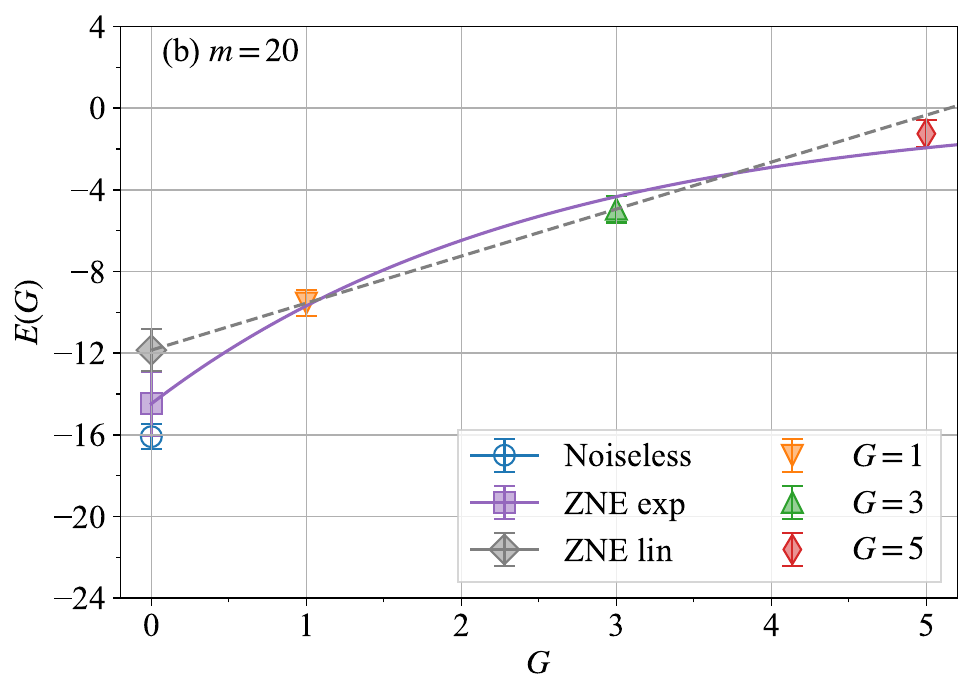}
\includegraphics[width=0.495\textwidth]{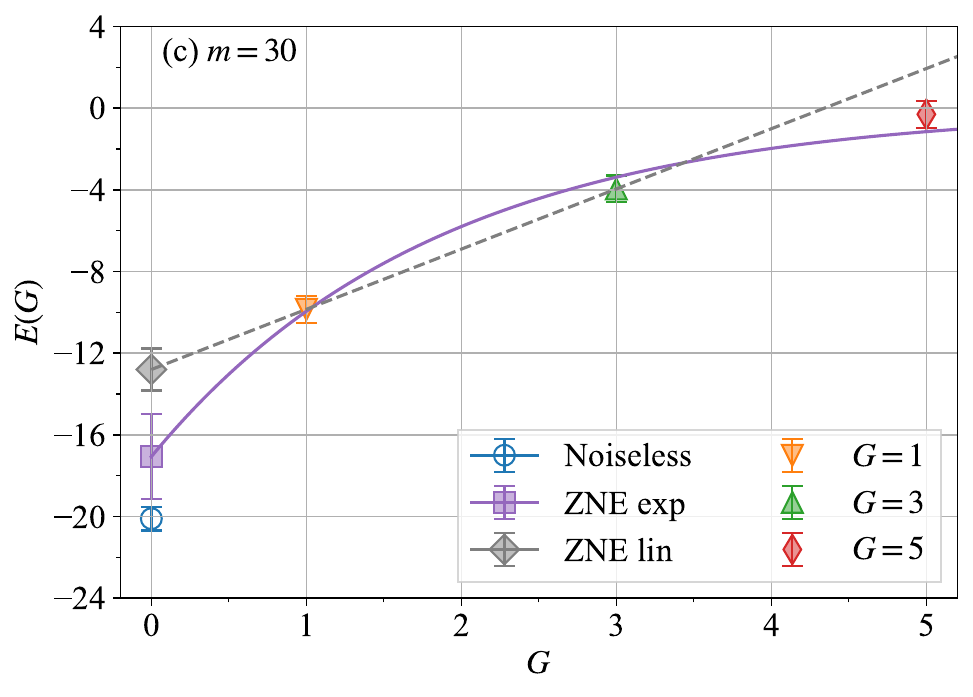}
\caption{
Same as Fig.~\ref{fig:ZNEdetails}, but for the $N=19$ transverse-field Ising model at time steps (a) $m=10$,  (b) $m=20$, and (c) $m=30$.
\label{fig:ZNEdetails19}}
\end{figure*}

In this Appendix, we present the details of the ZNE procedure for the $N=6$ and $N=19$ systems. 
Figure~\ref{fig:ZNEdetails} shows the linear and exponential fits applied to the energy expectation values obtained from the hardware experiments for the $N=6$ system as functions of the noise-scaling factor $G$ at time steps $m=5$, $10$, $15$, and $20$. 
Similarly, Fig.~\ref{fig:ZNEdetails19} displays the linear and exponential fits for the $N=19$ system at $m=10$, $20$, and $30$.

\bibliography{bib_dissipative_gs}

\end{document}